\documentclass[preprint,12pt,sort&compress]{elsarticle}

\usepackage{amssymb}

\usepackage{amsmath}
\usepackage{siunitx}
\usepackage{hyperref}
\usepackage{bm}
\usepackage{booktabs}
\usepackage{subcaption}

\def\equationautorefname~#1\null{%
  Equation~(#1)\null
}

\journal{Probabilistic Engineering Mechanics}

\begin{document}

    \newcommand{\Tinf}{T_{\infty}}
    \newcommand{\JTJ}{| \bm{J}^T \bm{J} |}
    \newcommand{\prob}[2][]{%
        \ifx\relax#1\relax
            p(\bm{#2})%
        \else
            p(\bm{#1}|\bm{#2})%
        \fi
    }

    \begin{frontmatter}
    
        

        \title{Bayesian estimation and uncertainty quantification of a temperature-dependent thermal conductivity}
        \author[ME,MCS,EAISI]{Rodrigo L. S. Silva\corref{cor1}}
        \ead{r.lima.de.souza.e.silva@tue.nl}
        \author[ME,EAISI]{Clemens Verhoosel}
        \author[MCS,EAISI]{Erik Quaeghebeur}
        \cortext[cor1]{Corresponding author}
        \affiliation[ME]{
            organization={Department of Mechanical Engineering, Eindhoven University of Technology},
            country={The Netherlands}
        }
        \affiliation[MCS]{
            organization={Department of Mathematics and Computer Science, Eindhoven University of Technology},
            country={The Netherlands}
        }
        \affiliation[EAISI]{
            organization={Eindhoven Artificial Intelligence Systems Institute, Eindhoven University of Technology},
            country={The Netherlands}
        }
        
        
        \begin{abstract}
            We consider the problem of estimating a temperature-dependent thermal conductivity model (curve) from temperature measurements.
            We apply a Bayesian estimation approach that takes into account measurement errors and limited prior information of system properties.
            The approach intertwines system simulation and Markov chain Monte Carlo (MCMC) sampling.
            We investigate the impact of assuming different model classes -- cubic polynomials and piecewise linear functions -- their parametrization, and different types of prior information -- ranging from uninformative to informative.
            Piecewise linear functions require more parameters (conductivity values) to be estimated than the four parameters (coefficients or conductivity values) needed for cubic polynomials.
            The former model class is more flexible, but the latter requires less MCMC samples.
            While parametrizing polynomials with coefficients may feel more natural, it turns out that parametrizing them using conductivity values is far more natural for the specification of prior information.
            Robust estimation is possible for all model classes and parametrizations, as long as the prior information is accurate or not too informative.
            Gaussian Markov random field priors are especially well-suited for piecewise linear functions.
        \end{abstract}
        
        
        
        \begin{keyword}
            
            

            Uncertainty quantification \sep
            Non-linear heat transfer \sep
            Temperature-dependent thermal conductivity \sep
            Markov chain Monte Carlo \sep
            Bayesian inference
        
        \end{keyword}
    
    \end{frontmatter}
    
    
    \section{Introduction}
    \label{section-introduction}

        The heat transfer analysis of engineering systems relies on the knowledge of different material properties, such as the thermal conductivity.
        The thermal conductivity is the proportionality constant in the Fourier's law, a constitutive model that relates the heat flux to the gradient of the temperature \cite{ozisik1993}.
        Often, the thermal conductivity is assumed to be constant.
        However, there are many engineering problems where the thermal conductivity varies significantly with the temperature, especially when dealing with large temperature ranges \cite{stelzer1987, ozisik1993, mota2010, ramos2022}.

        The measurement of the (temperature-dependent) conductivity is usually performed in laboratory experiments \cite{ozisik1993, aquino2006}, where it is possible to work with samples of the material of interest that assume simple geometries and have precise control over the experimental conditions.
        In situations where a controlled and representative lab experiment is not possible -- for example, in existing systems where it is infeasible to measure the relevant system properties with a high degree of certainty -- the measurement of a temperature-dependent conductivity can be challenging \cite{yang1999}.
        For such cases, techniques that can provide an estimation of this property are required \cite{ozisik1993, yang1999, aquino2006}.

        Different approaches for the estimation of temperature-dependent thermal constitutive models have been presented in the last few decades \cite{yang1999, alifanov1978, huang1991, huang1995, sawaf1995, chen1996, martin2000, lin2001, kim2002, aquino2006, zueco2006, mierzwiczak2011, czel2012, mohebbi2017}.
        These approaches are based on solving an inverse problem, in which the parameters of the thermal constitutive model are calibrated to match measurement data.
        Different techniques have been used for the solution of the inverse problem, including gradient-based algorithms \cite{yang1999, alifanov1978, huang1991, huang1995, sawaf1995, chen1996, kim2002, mohebbi2017}, neural networks \cite{aquino2006}, genetic algorithms \cite{czel2012} and others \cite{martin2000, lin2001, zueco2006, mierzwiczak2011}.
        The results obtained in these references convey that the proposed methods were able to provide a good estimation of the constitutive model parameters.
        Their focus has been on point estimates of the constitutive model parameters, not formally discussing the influence of uncertainties.
        However, the solution of the inverse problem can be extended beyond point estimates of the model parameters to also allow for the quantification of the uncertainties \cite{ozisik2018}.

        Bayesian techniques \cite{kaipio2006} provide a natural framework for uncertainty quantification, since the solution of the inverse problem is obtained through statistical inference \cite{ozisik2018, kaipio2006, silva2021design, silva2023design}.
        Different Bayesian frameworks for the estimation of constitutive models exist \cite{madireddy2015, madireddy2016, tao2021, aggarwal2023, yue2021, do2022, akintunde2019}.
        While Bayesian inference has been studied for linear transient heat problems \cite{orlande2008, naveira2010, naveira2011, gnanasekaran2011, lanzarone2014bayesian, berger2016bayesian, woo2022estimation}, its application to non-linear problems on account of temperature dependence of the conductivity has not been widely explored.
        Mota et al. \cite{mota2010} used Bayesian inference to estimate the coefficients of an exponential function that models the temperature dependence of the thermal conductivity of the material under analysis.
        Recently, Ramos et al. \cite{ramos2022} studied the Bayesian estimation of a temperature-dependent thermal conductivity model, with the assumption that the conductivity is constant on intervals of \SI{10}{\degreeCelsius}.
        Their results show that the extension to the non-linear case fundamentally alters the inverse problem.
        This happens because a strategy is required to model the thermal conductivity as a function of the temperature.
        The extension to the non-linear case then raises fundamental questions regarding, e.g., the influence of the strategy used to select the constitutive model and the considerations for describing prior information.

        In this work we present a detailed study of the effects of the use of different models for representing the temperature dependence of the thermal conductivity within a Bayesian framework.
        For this, we first introduce a prototypical case where we propose a (third degree) polynomial model for the temperature-dependent conductivity.
        For this model, two approaches to solve the inverse problem are proposed:
        \begin{enumerate}
            \item The first approach directly estimates the coefficients of the polynomial that models the conductivity as a function of the temperature.
            \item The second approach indirectly estimates the polynomial coefficients by first estimating conductivity values at specified temperatures and subsequently fitting the polynomial model.
        \end{enumerate}
        Although these approaches are closely related, we demonstrate that they lead to different choices in the set-up of the inverse problems, making it possible to clearly distinguish their advantages and disadvantages.

        A second conductivity model is proposed next, in which we use a piecewise linear representation of the thermal conductivity function.
        Our motivation to introduce this second model is to improve the inference procedure for situations in which it is not appropriate to use a third degree polynomial to model the conductivity as a function of the temperature.
        Similarly to the second approach of the polynomial model, we estimate conductivity values at specified temperatures.

        In the presentation of the two proposed conductivity models, we emphasize the ease-of-use of each model, as this is instrumental to their application to real-life problems.
        The insights obtained from the presented comparative analysis carries over to a broader class of problems, in particular to non-linear constitutive models with a structure similar to that encountered in the temperature-dependent heat transfer problem.

        Without loss of generality, we use the following methodology:
        \begin{enumerate}
            \item We consider a 1-D transient heat transfer problem.
            The proposed methodology, including the finite element approximation of the thermal problem, is also applicable to multidimensional cases (albeit with increased computational cost).
            \item A sensitivity analysis, which provides insights regarding the parameters that can be (simultaneously) estimated during the solution of the inverse problem, is an integral part of the presented uncertainty quantification methodology.
            \item To construct a fully controlled setting for the Bayesian inverse problems, synthetic data is generated using the forward problem in combination with a suitable noise model.
            \item We then define the likelihood function in terms of the noise model.
            The likelihood is combined with prior information to form a posterior distribution of the constitutive model parameters.
            \item This posterior is explored through the Metropolis-Hastings (MH) algorithm \cite{ozisik2018, kaipio2006}, a Markov chain Monte Carlo (MCMC) method \cite{gamerman2006, kaipio2006}.
        \end{enumerate}
        The considered approach not only provides an estimation of the thermal conductivity function, but also delivers uncertainty quantification using credible intervals for each of the estimated function values.

        This paper is structured as follows.
        In \autoref{section-transient-heat-conduction-problem} we present the transient heat conduction problem, describing the physical system, the corresponding mathematical formulation, and  details about the numerical method used to solve the forward problem.
        In \autoref{section-bayesian-estimation-and-uncertainty-quantification} we discuss the theory about Bayesian estimation and uncertainty quantification that is relevant for this work.
        Additionally, we also provide insights about the MH method used to estimate the posterior.
        In \autoref{section-application-of-the-bayesian-framework-to-the-heat-conduction-problem}, we show how we define and implement our Bayesian framework.
        We start by presenting the project design, followed by the definition of both the likelihood and the prior.
        In \autoref{section-results-and-discussion} we present the estimation and uncertainty quantification of the conductivity obtained with the considered models and their respective priors.
        Finally, in \autoref{section-conclusions} we discuss the conclusions of our work, which include the main advantages and disadvantages of the proposed models.

    \section{Transient heat conduction problem}
    \label{section-transient-heat-conduction-problem}

\begin{figure}
            \centering
            \includegraphics[scale=0.3]{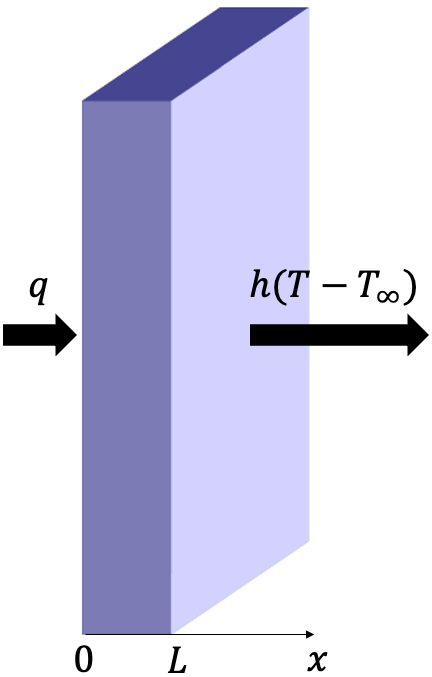}
            \hfil
            \includegraphics[width=0.38\linewidth]{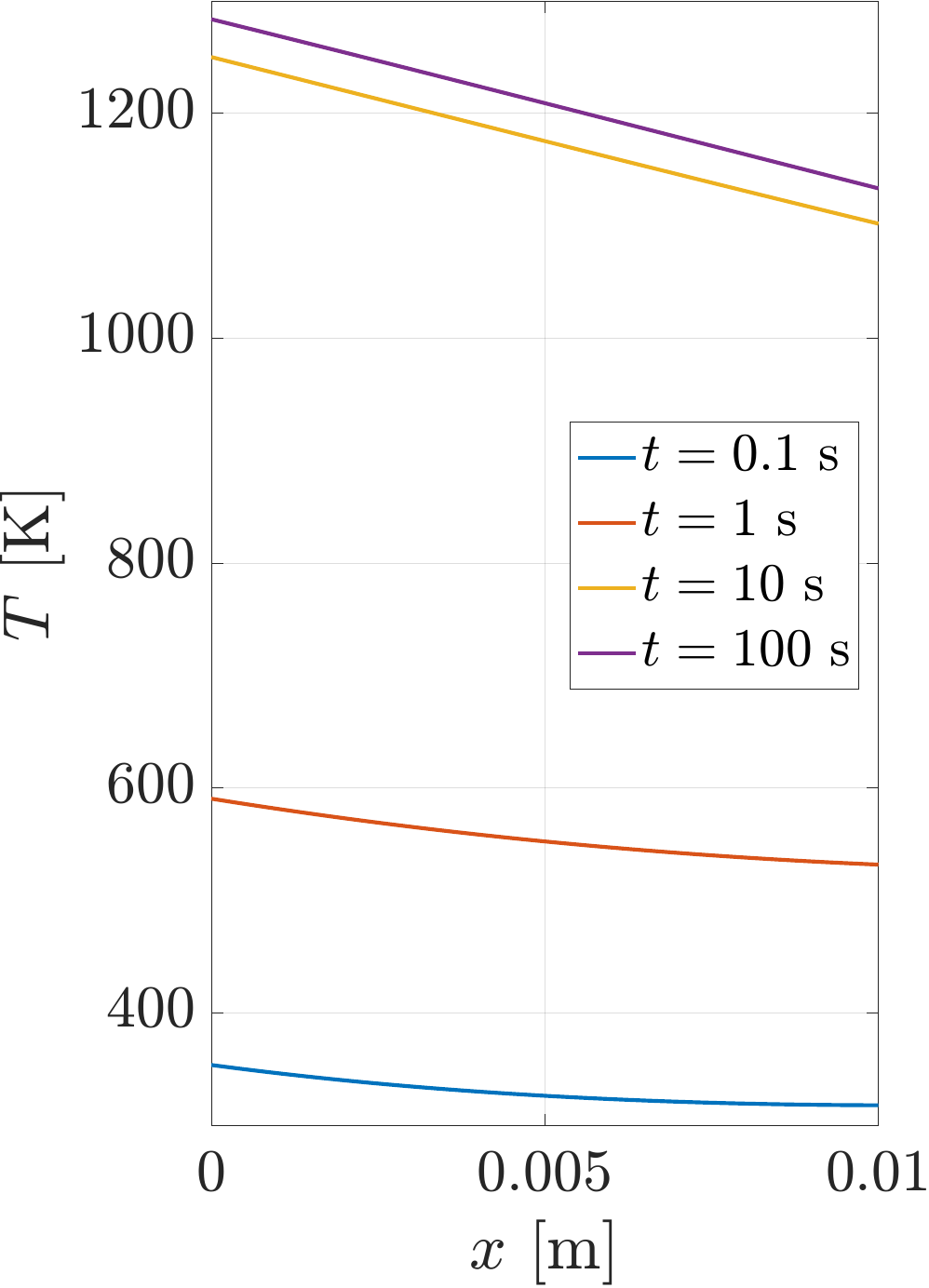}
            \caption{On the left, an illustration of the transient heat conduction problem. On the right, a typical modeling result, showing the temperature distribution along the slab at different times.}
            \label{domain}
        \end{figure}

        We consider a transient heat conduction problem in which a slab is heated on one end by a constant heat flux and cooled on the other end by a constant temperature ambient medium (\autoref{domain}).
        This problem setting results in a time-varying temperature profile along the slab. The direct problem is to obtain this time-varying temperature profile for the given boundary and initial conditions and material properties, specifically the thermal conductivity.

        In transient heat transfer scenarios, we can effectively utilize a limited number of (virtual) temperature sensors by obtaining multiple temperature measurements at various time instances.
        This approach is advantageous compared to steady-state scenarios, where it is conventional to gather data from a larger number of sensors distributed across diverse positions \cite{ozisik2018}.
        The transient temporal evolution of the temperature allows us to extract valuable information with just a few sensors, simplifying the data acquisition process and making it more practical.

        \subsection{Physical system}
        \label{subsection-physical-system}

            Our one-dimensional heat conduction problem consists of a slab of length $L = \SI{0.01}{m}$ (Figure \ref{domain}), with the coordinate system $x$ having its origin at the temperature inflow boundary.
            Initially, the temperate in the slab is $T_0 = \SI{300}{K}$.
            The slab is then heated by a constant heat flux $q = \SI{500e3}{W/m^2}$ at the left edge and cooled by a constant temperature medium at $T_\infty = \SI{300}{K}$ at the right edge, where a constant heat transfer coefficient $h = \SI{600}{W/m^2.K}$ is assumed.

            The relevant material properties of the slab are its specific heat $c_p = \SI{486}{J/kg.K}$, density $\rho = \SI{7870}{kg/m^3}$ and unknown temperature-dependent thermal conductivity $k(T)$.
            In transient heat conduction problems, also the temperature dependence of the density and specific heat can play an important role \cite{huang1991, huang1995, sawaf1995, mota2010, ramos2022}, but this work is focused on the constitutive model, so we consider these parameters to be temperature independent and known precisely.
            Hence, of the physical properties listed above, the thermal conductivity $k(T)$ has a special status, as this property needs to estimated as part of the solution to the inverse problem, which is presented in \autoref{section-application-of-the-bayesian-framework-to-the-heat-conduction-problem}.

        \subsection{Mathematical model}
        \label{subsection-mathematical-model}

            The physical system described above is modeled using the following mathematical formulation \cite{ozisik1993}:
            \begin{subequations}
            \label{equations-dimensional}
                \begin{align}
                    \rho c_p \dfrac{\partial T(x,t)}{\partial t} &= \dfrac{\partial}{\partial x} \left[ k(T) \dfrac{\partial T(x,t)}{\partial x} \right],
                    && 0 < x < L,
                    && t > 0;
                    \label{equation-dimensional-governing}
                    \\
                    k(T) \dfrac{\partial T(x,t)}{\partial x} &= -q,
                    && x = 0,
                    && t > 0;
                    \label{equation-dimensional-bc1}
                    \\
                    k(T) \dfrac{\partial T(x,t)}{\partial x} &= h[\Tinf - T(x,t)],
                    && x = L,
                    && t > 0;
                    \label{equation-dimensional-bc2}
                    \\
                    T(x,t) &= T_0,
                    && 0 \leq x \leq L,
                    && t = 0,
                    \label{equation-dimensional-ic}
                \end{align}
            \end{subequations}
            where $k(T) = k[T(x,t)]$.
            \autoref{equation-dimensional-governing} represents the energy balance.
            \autoref{equation-dimensional-bc1} and \eqref{equation-dimensional-bc2} describe the boundary conditions at $x = 0$ and $x = L$, respectively, and \autoref{equation-dimensional-ic} specifies the initial condition.

            We make the system of equations \eqref{equations-dimensional} dimensionless, to get a form that describes the essence of the problem and to obtain insights regarding the relevant dimensionless numbers.
            To obtain the dimensionless form, we introduce the dimensionless quantities $X = x / L$, $\tau = q t / \rho c_p T_0 L$, $\theta = T / T_0$, $\kappa = T_0 k / q L$ and $H = h T_0 / q$.
            These quantities respectively represent the dimensionless spatial coordinate, time, temperature, thermal conductivity and heat transfer coefficient.
            Additionally, we introduce the dimensionless ambient temperature $\theta_{\infty} = \Tinf / T_0$.
            Using these definitions, the dimensionless mathematical formulation reads:
            \begin{subequations}
            \label{equations-dimensionless}
                \begin{align}
                    \dfrac{\partial \theta(X, \tau)}{\partial \tau} &= \dfrac{\partial}{\partial X} \left[ \kappa(\theta) \dfrac{\partial \theta(X, \tau)}{\partial X} \right],
                    && 0 < X < 1,
                    && \tau > 0;
                    \label{equation-dimensionless-governing}
                    \\
                    \kappa(\theta) \dfrac{\partial \theta(X, \tau)}{\partial X} &= -1,
                    && X=0,
                    && \tau > 0;
                    \label{equation-dimensionless-bc1}
                    \\
                    \kappa(\theta) \dfrac{\partial \theta(X, \tau)}{\partial X} &= H [ \theta_{\infty} - \theta(X,\tau) ],
                    && X=1,
                    && \tau > 0;
                    \label{equation-dimensionless-bc2}
                    \\
                    \theta(X,\tau) &= 1,
                    && 0 \leq X \leq 1,
                    && \tau=0.
                    \label{equation-dimensionless-ic}
                \end{align}
            \end{subequations}
            This system of equations models the direct problem, the solution of which provides values of $\theta$ given $X$, $\tau$, $\kappa$, $H$ and $\theta_{\infty}$.

        \subsection{Numerical simulation}
        \label{subsection-numerical-simulation}

            We discretize the system of equations \eqref{equations-dimensionless} using linear finite elements \cite{lewis1996finite} in space and backward Euler \cite{butcher2016numerical} in time, for which the temperature field is approximated as
            \begin{equation}
                \theta(X,\tau) \approx \sum_{i=1}^{z} N_i(X) \theta_i(\tau) = \bm{N} \bm{\theta}, 
                \label{numerical}
            \end{equation}
            where $\bm{N} = [N_1(X), \dots, N_z(X)]$ is the (row) vector of $z$ linear basis functions constructed over a uniform mesh with element size $\Delta X$, and $\boldsymbol{\theta}(\tau) = [ \theta_1(\tau), \dots, \theta_z(\tau)]^T$ is the (column) vector of nodal temperatures.
            The time step size is fixed at $\Delta \tau$, such that $\tau^m = m \Delta \tau$, where $m = 1, 2, \dots$ is the time step index.
            We set the number of elements to 5, such that $\Delta X = 0.2$ and $z = 6$.
            Next to that, we set the dimensional time step as $\Delta t = 0.2$ seconds, which results approximately in $\Delta \tau = \num{8.7e-3}$.
            We select this value for the time step with the argument that it provides a good balance between the precision of the solution and computational time.
            Additionally, this time step is within the lower and upper bounds required to provide the stability of the numerical solution \cite{szabo2009discretization}.
            Finally, the temperature dependence of the thermal conductivity is treated explicitly, meaning that the values of the thermal conductivity $\kappa$ at time $\tau^{m+1}$ are based on the temperatures at time $\tau^m$.
            
            The nodal temperatures at time step $m + 1$, i.e., $\bm{\theta}^{m + 1} = \bm{\theta}(\tau^{m + 1})$, can be computed by solving the linear system
            \begin{equation}
                \bm{A}^m \bm{\theta}^{m + 1} = \bm{b}^m,
            \end{equation}        
            where $\bm{A} = (\Delta \tau)^{-1} \bm{C}+ \bm{K}^m + \bm{G}$ and $\bm{b}^m = (\Delta \tau)^{-1} \bm{C} \bm{\theta}^m + \bm{g}$.
            The terms $\bm{C}$ and $\bm{K}^m$ denote the capacity and conductance matrix, respectively.
            The matrix $\bm{G}$ and vector $\bm{g}$ are associated with the boundary conditions \cite{lewis1996finite}.
            $\bm{G}$ is a $z \times z$ matrix, in which all elements are zero except for the last element in the main diagonal, which is equal to $H$.
            The vector $\bm{g}$ has dimension $z$ and is defined as $\bm{g} = [1, 0, ..., 0, H \theta_{\infty}]^T$.
            Additionally, the matrices $\bm{C}$ and $\bm{K}^m$ are defined as
            \begin{subequations}
            \begin{align}
                \bm{C} &= \int_0^1 \bm{N}^T\bm{N} \, dX,
                \\
                \bm{K}^m &=  \int_0^1 \kappa ( \bm{\theta}^m ) \dfrac{d\bm{N}^T}{dX} \dfrac{d\bm{N}}{dX} \, dX.
                \label{conductance_matrix}
            \end{align}    
            \end{subequations}
            In \autoref{conductance_matrix}, the value of $\kappa$ used for a specific element is assumed constant and is evaluated by using the mean temperature of the nodes that constitute this element.

    \section{Bayesian estimation and uncertainty quantification}
    \label{section-bayesian-estimation-and-uncertainty-quantification}

        In this section we introduce the key concepts of the Bayesian framework required to apply it to the non-linear heat conduction problem introduced above.

        \subsection{Bayesian modeling}
        \label{subsection-bayesian-modeling}

            In many practical applications, prior information regarding the parameters to be estimated is available. This information can be obtained, for example, from previous experiments, the literature, or experts' opinions.
            Bayesian frameworks are useful for such cases, since prior information can be formally taken in account during the estimation of the parameters of interest.
            
            The estimation of parameters within the Bayesian framework can be summarized in the following steps \cite{ozisik2018, kaipio2006, silva2023design}:
            We start by defining $\bm{P}$ as the vector with the parameters to be estimated.
            The number of parameters -- and therefore, the number of elements in $\bm{P}$ -- is denoted here by $N$.
            Based on all information available for $\bm{P}$, we select a prior distribution $\prob{P}$, which represents the statistical model of the information available for the parameters.
            Then we select the likelihood function $\prob[D]{P}$.
            This likelihood captures how a data vector $\bm{D}$ is related to the parameters $\bm{P}$.
            Finally, we explore the posterior distribution $\prob[P]{D}$, which is the conditional probability distribution of the unknown parameters given the data.
            
            In this way, Bayesian estimation amounts to obtaining the posterior distribution.
            This posterior is obtained using Bayes' theorem \cite{ozisik2018, kaipio2006}, which states that
            \begin{equation}
                \prob[P]{D} = \dfrac{\prob{P} \prob[D]{P}}{\prob{D}}.
                \label{bayes}
            \end{equation}
            In the equation above, $\prob{D}$ is called the evidence and plays the role of a normalizing constant.
            The computation of the evidence $\prob{D}$ is usually not needed for practical calculations \cite{ozisik2018, kaipio2006}, and so Bayes' theorem reduces to
            \begin{equation}
                \prob[P]{D} \propto \prob{P} \prob[D]{P}.
                \label{bayes-prop}
            \end{equation}
            \autoref{bayes-prop} shows that the posterior $\prob[P]{D}$ depends not only on the likelihood function $\prob[D]{P}$, but also on the prior distribution $\prob{P}$.
            Hence, even if the prior information is only qualitatively available, it must be mathematically modeled as a statistical distribution \cite{ozisik2018}.
            
            The use of the Bayesian framework for the solution of inverse problems can provide not only point estimates, but also statistics of the posterior distribution that allow for the quantification of the related uncertainties.
            However, the direct computation of these statistics typically requires numerical integration, which is often impractical from a computational effort point of view \cite{ozisik2018, kaipio2006}.
            For such cases, stochastic simulation with Markov chain Monte Carlo (MCMC) methods \cite{gamerman2006, ozisik2018} can provide an indirect computational approach.

        \subsection{Stochastic simulation with Markov chain Monte Carlo methods}
        \label{subsection-stochastic-simulation-with-Markov-chain-Monte-Carlo-methods}

            In MCMC methods, samples of the posterior distribution are generated by stochastic simulation.
            Then inference on the posterior distribution -- i.e, deriving statistics and expectations -- is performed through inference on these samples \cite{ozisik2018, kaipio2006}.
            MCMC methods can in this way approximate the posterior.
            
            A Markov chain is a stochastic process in which, given the present state, past and future states are independent \cite{gamerman2006}.
            If a Markov chain, independently of its initial distribution, reaches a stage that can be represented by a specific distribution $\lambda$, and retains this distribution for all subsequent stages, we say that $\lambda$ is the limit distribution of the chain \cite{gamerman2006}.
            For the Bayesian approach, the Markov chain is constructed in such a way that its limit distribution coincides with the posterior.
            The literature on Markov chains \cite[\textit{e.g.},][]{gamerman2006} contains more details for the interested reader.
            
            We use the Metropolis-Hastings (MH) algorithm \cite{ozisik2018, kaipio2006, gamerman2006} as our MCMC method.
            We selected this algorithm because it is easy to implement and widely used for the solution of inverse problems \cite{ozisik2018, kaipio2006}.
            The MH algorithm starts with the definition of a initial guess $\bm{P}^0$ for the vector of parameters.
            A new candidate is then sampled from a proposal distribution.
            We define this proposal as a multivariate normal distribution, with its mean centered at the current state and a specific covariance matrix.
            We obtain this covariance matrix from the adaptive MCMC algorithm proposed by Haario et al. \cite{haario2001adaptive}.
            The new candidate is then either accepted or rejected according to a probabilistic criterion.
            This procedure is repeated in order to generate the Markov chain $[\bm{P}^1, \bm{P}^2, \dots, \bm{P}^R]$, where $R$ is the number of steps of the algorithm.
            
            One usually notices an initial sequence $[\bm{P}^1, \bm{P}^2, \dots, \bm{P}^r]$, $1 < r < R$, which contains all samples before reaching equilibrium. 
            This sequence is called the burn-in period \cite{ozisik2018, kaipio2006, gamerman2006}.
            For all Markov chains shown in this work, the burn-in period is determined via visual inspection.
            If the chain $[\bm{P}^{r+1}, \bm{P}^{r+2}, ..., \bm{P}^R]$ satisfies a convergence criterion, then it is used to represent samples from the limit distribution, and therefore from the posterior.
            The convergence criterion we use in this work is the one proposed by Geweke \cite{geweke1991evaluating}.
            This criterion evaluates the mean of the samples of the first 10\% and of the last 50\% of the states in the chain $[\bm{P}^{r+1}, \bm{P}^{r+2}, ..., \bm{P}^R]$.
            We respectively denote these means as $m_{10}$ and $m_{50}$.
            If the difference between these means is small, then the convergence criterion is satisfied \cite{ozisik2018, geweke1991evaluating}.
            We implement this criterion by checking the values of $| (m_{10} - m_{50}) / m_{10} |$ and $| (m_{10} - m_{50}) / m_{50} |$.
            If both values are smaller than or equal to $10^{-2}$, then it is satisfied.
            For all Markov chains shown in this paper, the samples used to represent each limit distribution (and therefore, the posterior) are selected only if the convergence criterion mentioned above is satisfied.

    \section{Application of the Bayesian framework to the heat conduction problem}
    \label{section-application-of-the-bayesian-framework-to-the-heat-conduction-problem}

        \subsection{Definition of simulated temperature measurements}
        \label{subsection-definition-of-simulated-temperature-measurements}

            As discussed in \autoref{section-bayesian-estimation-and-uncertainty-quantification}, the Bayesian framework used in this paper requires data obtained from the system under analysis.
            In practical heat conduction experiments, these data are usually temperature measurements, which can be obtained, for example, from temperature sensors or infrared cameras.
            We focus here on cases involving temperature sensors, where transient temperatures are measured at fixed locations.
            
            We herein use simulated temperature measurements as the data for the Bayesian framework.
            These simulated measurements are obtained as follows:
            We first define the positions where the temperature will be simulated.
            Then we solve the direct problem represented by the system of equations \eqref{equations-dimensionless} for these positions and various time instances.
            Finally, the simulated measurements are defined by adding random noise to these solutions.
            The random noise is assumed normal and uncorrelated, with its mean and standard deviation specified below.
            
            We consider simulated measurements obtained from sensors located at both edges of the slab (i.e., $X = 0$ and $X = 1$).
            These measurements are acquired for a total of 10 minutes, with the same time step $\Delta t = 0.2$~s that is used to discretize the time domain for the numerical simulation.
            This results in $M = 3000$ measurements for each sensor.
            We then define $\bm{T}_0$ as the vector with the $M$ values of transient temperatures obtained from the solution of the direct problem at $X = 0$.
            The definition of $\bm{T}_1$, which is obtained assuming $X = 1$, is analogous.
            Next, we define the vector $\bm{T} = [\bm{T}_0, \bm{T}_1]^T$, with dimension $U = 2 M$.
            Additionally, we denote the vector with the random noise as $\bm{E}$.
            It has dimension $U$, and each element $E_j$, $j = 1, \dots, U$, is a random number obtained from a normal distribution with zero mean and standard deviation $\sigma_j$.
            Next, we set $\sigma_j = T_j / 100$, where $T_j$ represents an element in $\bm{T}$.
            Finally, the vector $\bm{D}$, which contains all the $U$ simulated measurements that are used in the Bayesian analysis, is defined as $\bm{D} = \bm{T} + \bm{E}$.
            
            In order to solve the direct problem and generate the simulated temperature measurements, we need to choose a ‘ground truth’ model to represent the conductivity as a function of the temperature.
            The model we use is the one reported by Aquino and Brigham \cite{aquino2006}, which represents the temperature dependence of the thermal conductivity by a third degree polynomial as $\kappa(\theta) = \sum_{n=1}^4 C_n \theta^{4 - n}$, where $C_1 = 0.0810$, $C_2 = -0.4860$, $C_3 = 0.0918$ and $C_4 = 4.2060$.

        \subsection{Project design using sensitivity analysis}
        \label{subsection-project-design-using-sensitivity-analysis}

            We conduct the project design through a sensitivity analysis \cite{ozisik2018} on the parameters of interest.
            This allows us to obtain insight about which parameters can be simultaneously estimated \cite{ozisik2018}.
            Hence, the sensitivity analysis is performed before solving the inverse problem.
            
            We start the sensitivity analysis by defining the sensitivity coefficients.
            Since we want to understand how changes in the vector of parameters results in changes in the estimated temperatures, it is convenient to write $\theta$ as a function of $\bm{P}$.
            Hence, a value of the temperature obtained at position $X$ and time $\tau$, given the vector of parameters $\bm{P}$, is denoted here as $\theta(X, \tau, \bm{P})$.
            If we consider the temperature sensor located at $X=0$, the sensitivity coefficient $J_{0,mn}$ is defined as the first derivative of $\theta$, evaluated at $X =0$ and time $\tau_m$, with respect to the parameter $P_n$, that is,
            \begin{equation}
                J_{0,mn}(\bm{P}) = \dfrac{\partial \theta(0, \tau_m, \bm{P})}{\partial P_n}.
                \label{Jmn}
            \end{equation}
            Each sensitivity coefficient $J_{0,mn}$, for $m = 1, \dots, M$ and $n = 1, \dots, N$, is an element of the sensitivity matrix $\bm{J}_0$ \cite{ozisik2018}, which has dimension $M \times N$.
            
            For simplicity, we now consider a case in which we only have temperature values obtained at $X = 0$.
            A small value of the magnitude of $J_{0,mn}$ indicates that large changes of $P_n$ yield small changes in $\theta(0, \tau_m, \bm{P})$.
            The estimation of $P_n$ is difficult in such case, because basically the same value of $\theta(0, \tau_m, \bm{P})$ would be obtained for a wide range of values of $P_n$.
            In fact, when all the sensitivity coefficients are small, we have $\JTJ \approx 0$ and the inverse problem is ill-conditioned \cite{ozisik2018}.
            It is also possible to demonstrate that $\JTJ = 0$ if a column in $\bm{J}$ can be expressed as a linear combination of the other columns \cite{ozisik2018}.
            Hence, it is desired to have linearly independent sensitivity coefficients with large magnitudes.
            This allows us to achieve accurate estimation and uncertainty quantification of the parameters of interest.
            
            The derivative in \autoref{Jmn} is discretized in this work with the central finite difference method \cite{ozisik2018}.
            That means we approximate $J_{0,mn}$ as follows \cite{ozisik2018}:
            \begin{equation}
                J_{0,mn}(\bm{P}) \approx \dfrac{\theta(0, \tau_m, \bm{P^+}) - \theta(0, \tau_m, \bm{P^-})}{2 \epsilon P_n},
                \label{J_mn-FD}
            \end{equation}
            where
            \begin{align}
                \bm{P^+} &= (P_1, \dots, P_n + \epsilon P_n, \dots, P_N)^T
                \\
                \bm{P^-} &= (P_1, \dots, P_n - \epsilon P_n, \dots, P_N)^T
            \end{align}
            and $0 < \epsilon \ll 1$ (we set $\epsilon = \num{e-5}$).
            We notice from Equations \eqref{Jmn} and \eqref{J_mn-FD} that, since the derivative is locally evaluated, we need to specify a value for $P_n$.
            The true sensitivity coefficient is obtained when the true value of $P_n$ is provided.
            Since $P_n$ is a parameter to be estimated as part of the solution of the inverse problem, we do not know its true value.
            Hence, a reference value must be provided in order to evaluate the sensitivity coefficient defined in \autoref{J_mn-FD}.
            If this reference value is sufficiently close to the true one, it is possible to obtain a good estimate of the sensitivity coefficient.
            In the context of Bayesian inference, prior information regarding the parameters to be estimated is also useful for project design, since it allows us to obtain better estimates of the sensitivity coefficients.
            
            The definitions of the sensitivity coefficient $J_{1,mn}$ and the sensitivity matrix $\bm{J}_1$, which are evaluated at $X = 1$, as well as the finite difference approximation, are analogous to those presented above.
            We can then define $\bm{J} = [\bm{J}_0, \bm{J}_1]^T$, which has dimension $U \times N$ and represents the vertical concatenation of $\bm{J}_0$ and $\bm{J}_1$.
            This allows us to combine the information of both sensors located at $X = 0$ and $X = 1$.

        \subsection{The likelihood function}
        \label{subsection-the-likelihood-function}

            In \autoref{subsection-definition-of-simulated-temperature-measurements} we construct the components $D_j$, $j = 1, \dots, U$, from the data vector $\bm{D}$ as coming from a normal distribution.
            This distribution has mean $T_j$, which is an element from the vector with the solution $\bm{T}$ of the direct problem, and standard deviation $\sigma_j$.
            This expresses that we work under the assumption that the data generation process is Gaussian.
            Therefore, it is reasonable to take the likelihood function to be a multivariate normal distribution.
            This distribution has a mean vector equal to the solution of the direct problem $\bm{T}(\bm{P})$ and a covariance matrix $\bm{W}$.
            This covariance matrix $\bm{W}$ is a $U \times U$ diagonal matrix, with elements $W_{jj} = \sigma_j^2$.
            So we can express the likelihood function as
            \begin{equation}
                \prob[D]{P} \propto \exp \left[ -\dfrac{1}{2} \bm{E(P)}^T \bm{W}^{-1} \bm{E(P)} \right],
            \end{equation}
            where $\bm{E}(\bm{P})=\bm{D}-\bm{T}(\bm{P})$ is the measurement error implied by the current estimate $\bm{P}$.

        \subsection{Temperature-dependent thermal conductivity models}
        \label{subsection-temperature-dependent-thermal-conductivity-models}

            Two different model classes for the thermal conductivity as a function of the temperature are considered in this paper.
            We choose these two classes because they lead to different choices in the set-up of the inverse problem (e.g., the selection of the prior), which affects the estimation of the posterior.
            The selected classes are clearly distinguishable in terms of their advantages and disadvantages.
            We assume that both classes are valid for values of $\theta$ between $\theta_{\text{min}}$ and $\theta_{\text{max}}$, which are respectively the minimum and maximum measured temperatures.
            These values also represent, respectively, the lower and upper bounds of the dimensionless temperature range of interest.
            
            Our first class of models for representing the conductivity as a function of the temperature are the third-degree polynomials.
            We choose these due to their smooth behavior, which can be appropriate to represent a thermal conductivity as a function of the temperature \cite{aquino2006}.
            Furthermore, as discussed in \autoref{subsection-definition-of-simulated-temperature-measurements}, we use a third degree polynomial to represent the conductivity as a function of the temperature to generate our data, so we decide to test the framework by first selecting from the same model class as the ground truth model.
            
\begin{figure}
                \centering
                \begin{subfigure}[t]{0.45\linewidth}
                    \vskip 0pt
                    \includegraphics[width=\linewidth]{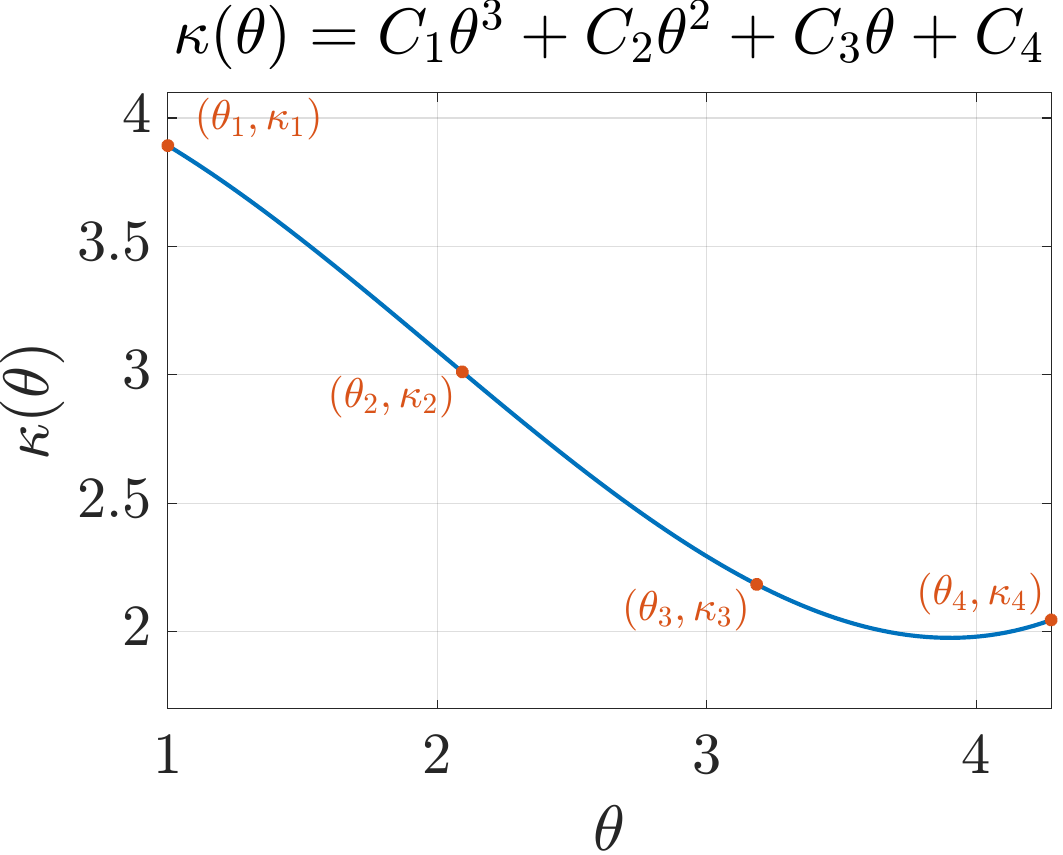}
                    \caption{A third-degree polynomial.
                    This polynomial model can either be parametrized directly by the coefficient, $C_n$, or indirectly through the points $(\theta_n,\kappa_n)$.}
                    \label{figure-polynomial-model}
                \end{subfigure}
                \hfil
                \begin{subfigure}[t]{0.45\linewidth}
                    \vskip 0pt
                    \includegraphics[width=\linewidth]{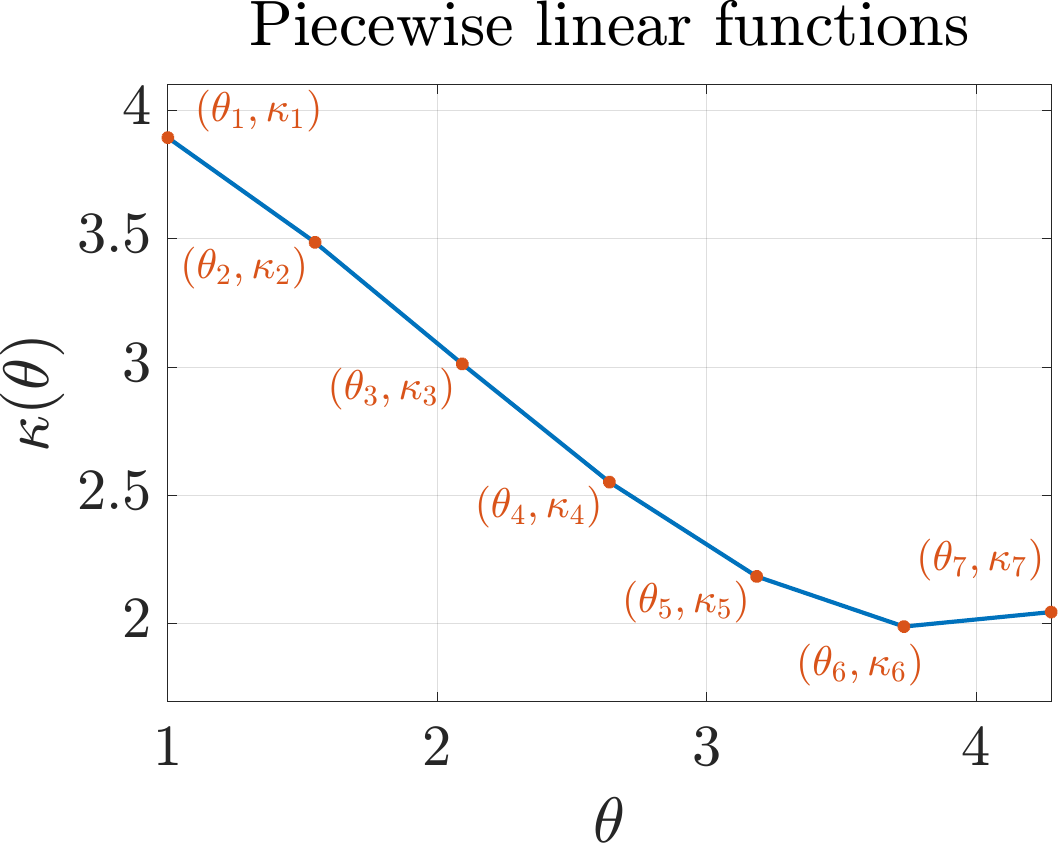}
                    \caption{Piecewise linear functions.
                    This model is parametrized through the points $(\theta_n,\kappa_n)$.}
                    \label{figure-piecewise-model}
                \end{subfigure}
                \caption{The temperature-dependent conductivity function $\kappa(\theta)$ represented by two models.}
                \label{}
            \end{figure}

            We consider two different parametrizations for the polynomial model:
            The first parametrization, discussed in \autoref{subsubsection-coefficients-parametrization}, considers directly estimating the coefficients of the polynomial.
            The second parametrization, discussed in \autoref{subsubsection-conductivity-values-parametrization}, indirectly estimates the coefficients of the polynomial by first estimating values of the conductivity $\kappa_n = \kappa(\theta_n)$, $n = 1, \dots, 4$, where $\theta_n$ are equally spaced within the temperature range of interest, and subsequently fitting the polynomial to these values.
            We use these two parametrizations to illustrate how they alter the selection of the prior.
            The polynomial model and its parametrizations are illustrated in \autoref{figure-polynomial-model}.

            Our second class of models are the piecewise linear functions.
            Our motivation to use this class is to obtain better estimates in cases where it is not convenient to assume a predetermined functional form to model the conductivity as a function of the temperature (such as a third-degree polynomial).
            In order to use this class, we discretize the function $\kappa(\theta)$ with a sufficiently large number of points, and we aim at estimating these discretized values of the conductivity.
            The piecewise linear functions model is illustrated in \autoref{figure-piecewise-model}.

        \subsection{Third-degree polynomial model class}
        \label{subsection-third-degree-polynomial-model-class}

            We assume here that the thermal conductivity is modeled as a function of the temperature by a third-degree polynomial.
            This implies in $\kappa(\theta) = \sum_{n=1}^4 C_n \theta^{4 - n}$, where the coefficients $C_n$ are real and dimensionless.

            \subsubsection{Coefficients parametrization}
            \label{subsubsection-coefficients-parametrization}

                In this section the estimation of $\kappa$ is obtained via the estimation of the polynomial coefficients $C_n$.
                The vector of parameters is defined as $\bm{P} = [C_1, C_2, C_3, C_4]^T$.
                These coefficients do not have a clear physical meaning and therefore it can be difficult to select an informative prior.
                Hence, we select here a non-informative improper uniform prior \cite{kaipio2006, ozisik2018} for $\bm{P}$.
                Improper priors do not integrate to a finite value, and are often used in situations where we have limited prior knowledge or want to express non-informative beliefs \cite{kaipio2006, ozisik2018}.
                Nevertheless, improper priors still allow us to obtain proper posteriors \cite{kaipio2006, ozisik2018}.
                
                According to the second law of thermodynamics, the thermal conductivity must always be positive \cite{ozisik1993}.
                This information can be taken in account by truncating the prior distribution to exclude sets of coefficients $C_n$ that result in at least one non-positive value of $\kappa$ for a given $\theta$ between $\theta_{\text{min}}$ and $\theta_{\text{max}}$.
                The improper truncated prior distribution is then given by
                \begin{equation}
                    \prob{P} = 
                        \begin{cases}
                        1, \text{ if } \bm{P} \in \bm{\Phi}, \\
                        0, \text{ elsewhere},
                        \end{cases}
                    \label{uniformPrior}
                \end{equation}
                where $\bm{\Phi}$ represents the set of coefficients $C_n$ that result in positive values of $\kappa(\theta)$ for all $\theta$ between $\theta_{\text{min}}$ and $\theta_{\text{max}}$.
                
                The strategy used to verify whether $\bm{P} \in \bm{\Phi}$ is as follows:
                First, we check if $\kappa(\theta_{\text{min}})$ and $\kappa(\theta_{\text{max}})$ are positive.
                Then, we calculate the extrema using ${\rm d} \kappa(\theta) / {\rm d} \theta = 0$.
                For each extremum $\theta_e$, if it is real and $\theta_{\text{min}} < \theta_e < \theta_{\text{max}}$, we check if $\kappa(\theta_e)$ is positive.
                If these criteria are satisfied, then $\bm{P} \in \bm{\Phi}$.

            \subsubsection{Conductivity values parametrization}
            \label{subsubsection-conductivity-values-parametrization}

                In this section the estimation of the polynomial coefficients is performed by first estimating four values of the conductivity $\kappa_n = \kappa(\theta_n)$, $n = 1, 2, 3, 4$.
                The temperatures $\theta_n$ are equally spaced in the range between $\theta_{\text{min}}$ and $\theta_{\text{max}}$, with $\theta_1 = \theta_{\text{min}}$ and $\theta_4 = \theta_{\text{max}}$.
                We select four conductivities because that allows us to have four equations where the conductivities $\kappa_n$ are modeled as a function of the respective temperatures $\theta_n$ by a third-degree polynomial.
                These equations can be used to represent a linear system, and its solution provides us the values of the four coefficients $C_n$.
                The coefficients are then used to represent $\kappa$ as a function of $\theta$ in the temperature domain of interest.
                
                In this case the vector of parameters is equal to $\bm{P}=[\kappa_1,\kappa_2,\kappa_3,\kappa_4]^T$.
                Since the elements in $\bm{P}$ now have a clear physical meaning, it is easier to select the prior when compared to the parametrization based on the polynomial coefficients.
                For instance, values of the thermal conductivity, as a function of the temperature, for different materials can be found in the literature \cite{vosteen2003influence, osman2001temperature, shanks1963thermal}.
                These values can then be used as a reference for the definition of the priors regarding the conductivities $\kappa_n$.
                
                We analyze two different priors for $\bm{P}$: uniform and normal.
                Both priors are truncated, as specified below.
                The uniform is again assumed improper and is also defined by \autoref{uniformPrior}.
                For the normal prior, we denote its covariance matrix $\bm{V}$, with dimension $N \times N$.
                Additionally, the conductivities $\kappa_n$ are assumed independent, so the covariance matrix is diagonal with elements $V_{nn} = \sigma_n^2$, where $\sigma_n^2$ are the variances of $\kappa_n$.
                Hence, the normal prior is given by
                \begin{equation}
                    \prob{P} \propto
                    \begin{cases}
                        \exp \left[ -\dfrac{1}{2} ( \bm{P} - \bm{\mu} )^T \bm{V}^{-1} ( \bm{P} - \bm{\mu} ) \right], \text{ if } \bm{P} \in \bm{\Phi},
                        \\
                        0, \text{ elsewhere}.
                    \end{cases}
                \end{equation}
                The same strategy to verify whether $\bm{P} \in \bm{\Phi}$ as used with the coefficients parametrization is used here.
                The only difference is that, since now $\bm{P}$ contains values of $\kappa_n$, it is first necessary to calculate the corresponding coefficients of the third-degree polynomial by solving the linear system mentioned above.

        \subsection{Piecewise linear functions model}
        \label{subsection-piecewise-linear-functions-model}

            As a second model class we consider the temperature-dependent conductivity to be represented by a piecewise linear function.
            Our motivation is to analyze cases where it is not convenient to assume a predetermined functional form for the conductivity as a function of the temperature (\textit{e.g.}, a third-degree polynomial).
            We consider here an initial procedure similar to the one presented for the parametrization based on the conductivity values of the polynomial model:
            We first discretize the temperature domain with $N$ points $\theta_n$.
            Then we will estimate the values of the conductivity $\kappa_n = \kappa(\theta_n)$.
            These values represent the discretization of the function $\kappa(\theta)$.
            
            The number of points (and parameters to be estimated) $N$ must be sufficiently large in order to provide a good approximation of the function $\kappa(\theta)$.
            The increase in the number of parameters usually results in the increase of the number of samples of a Markov chain in order to reach the limit distribution \cite{kaipio2006, ozisik2018}.
            We consider here $N = 100$, with the argument that this value provides a good balance between the approximation of the function $\kappa(\theta)$ and computational time.
            Additionally, the increase in the number of parameters also alters the way we select our prior distribution.
            For example, the use of a normal distribution with a diagonal covariance matrix is usually not appropriate.
            The reason for this is that when the parameters are discrete function values, they cannot be assumed to be independent \cite{ozisik2018}.
            There is a correlation between these values, which must be taken in account while formulating the prior.
            The dependence regarding the elements in $\bm{P}$ is even more evident among neighboring parameters.
            
            In such a case, the prior information is better modeled by a Gaussian Markov random field (GMRF) \cite{ozisik2018, gamerman2006}.
            This prior is also referred to as a Gaussian smoothness prior \cite{ozisik2018, kaipio2006, mota2010}.
            We initially assume that $\kappa_{i+1} = \kappa_i + \delta_i$, $i = 1, 2, 3, ..., N - 1$, where $\delta_i$ is a normal uncorrelated process with zero mean and variance $\gamma^2$ \cite{mota2010, kaipio2006}.
            The prior distribution can then be written as \cite{ozisik2018, kaipio2006, gamerman2006, mota2010}
            \begin{equation}
                \prob{P} \propto
                \begin{cases}
                    \exp \left[ -\dfrac{1}{2 \gamma^2} (\bm{P} - \bar{\bm{P}})^T \bm{Z}^T \bm{Z} (\bm{P} - \bar{\bm{P}}) \right], \text{ if } \text{min}(\bm{P}) > 0,
                    \\
                    0, \text{ elsewhere,}
                \end{cases}
                \label{gmrf-initial}
            \end{equation}
            where $\text{min}(\bm{P})$ denotes the minimum value in $\bm{P}$.
            We impose the condition $\text{min}(\bm{P}) > 0$ to ensure that the piecewise linear functions will always result in positive values of $\kappa(\theta)$.
            Next to that, $\bar{\bm{P}}$ is a reference value for $\bm{P}$ and
            \begin{equation}
                \bm{Z} =
                    \begin{bmatrix}
                        -1 & 1 & 0 & 0 & \dots & 0 \\
                        0 & -1 & 1 & 0 & \dots & 0 \\
                        \vdots & \vdots & \vdots & \vdots & \vdots & \vdots \\
                        0 & \dots & 0 & 0 & -1 & 1
                    \end{bmatrix}_{(N - 1) \times N}.
            \end{equation}
            
            We rewrite \autoref{gmrf-initial} as follows:
            First, we define $\bm{Q(P)} = \bm{Z} \bm{P}$.
            We notice that an element $Q_i$ from vector $\bm{Q}$ represents the difference of two consecutive elements in $\bm{P}$, that is, $Q_i = P_{i+1} - P_i$, $i = 1, 2, 3, ..., N - 1$.
            Next, we define $\bar{\bm{Q}} = \bm{Z} \bar{\bm{P}}$.
            The representation of the elements in $\bar{\bm{Q}}$ is analogous to the one in $\bm{Q}$.
            We then have:
            \begin{equation}
                \prob{P} \propto
                \begin{cases}
                    \exp \left\{ -\dfrac{1}{2 \gamma^2} [\bm{Q}(\bm{P}) - \bar{\bm{Q}}]^T [\bm{Q}(\bm{P}) - \bar{\bm{Q}}] \right\}, \text{ if } \text{min}(\bm{P}) > 0,
                    \\
                    0, \text{ elsewhere.}
                \end{cases}
                \label{gmrf-intermediate}
            \end{equation}
            The equation above shows us that the GMRF prior models the differences of consecutive neighbours in $\bm{P}$ as a multivariate normal distribution.
            This distribution has mean vector $\bar{\bm{Q}}$ and covariance matrix $\gamma^2 \bm{I}$, where $\bm{I}$ is the $N \times N$ identity matrix.
            
            Note that, according to the relation $\bar{\bm{Q}} = \bm{Z} \bar{\bm{P}}$, different values of $\bar{\bm{P}}$ can result in the same $\bar{\bm{Q}}$.
            Consider, as an example, $\bar{\bm{P}}^* = \bar{\bm{P}} + \bm{F}$, where $\bm{F}$ is a vector with all elements equal to a real constant.
            Then $\bm{Z} \bar{\bm{P}}^* = \bm{Z} \bar{\bm{P}} + \bm{Z} \bm{F}$.
            Since $\bm{Z} \bm{F} = \bm{0}$, where $\bm{0}$ is a zero vector, $\bm{Z} \bar{\bm{P}}^* = \bm{Z} \bar{\bm{P}}$.
            This means that adding a constant value to $\bar{\bm{P}}$ does not alter the result of $\bar{\bm{Q}}$.
            Hence, the GMRF prior only constrains $\bm{P}$ up to a constant, and it is improper with relation to the mean.

    \section{Results and discussion}
    \label{section-results-and-discussion}

        In this section we present the results obtained with the third-degree polynomial and piecewise linear functions models.
        We also discuss how these two models and the corresponding selection of the prior alter the estimation and uncertainty quantification of the conductivity. For all cases discussed below, the proposal distribution covariance matrices are obtained from the adaptive MCMC. These covariances resulted in acceptance ratios between 25\% and 35\% for all cases. All simulations were performed on personal computer with an Intel Core i7-10750H processor and 16 GB RAM.

        \subsection{Third-degree polynomial model}
        \label{subsection-results-third-degree-polynomial-model}

            In \autoref{subsection-third-degree-polynomial-model-class} we presented two different parametrizations for the case where the conductivity is modeled as a function of the temperature by a third-degree polynomial. In \autoref{subsubsection-results-conductivity-values-parametrization} we will first discuss the results for the parametrization based on conductivity values. In \autoref{subsubsection-results-coeficient-parametrization} we will then discuss the parametrization based on the polynomial coefficient.
            In this section we will also present a comparison of the two parametrizations.

            For all simulation in this section we consider a total of \num{20000} samples for the adaptive MCMC algorithm.
            This was sufficient to provide the covariance matrix used to generate the proposals in the MH algorithm.
            For all cases the computational time required to generate the samples with the adaptive MCMC was approximately 10 minutes. For the MH algorithm we consider a total of \num{5000} samples. The burn-in period is defined as the first \num{1000} samples, based on visual inspection of the chains. The computation time to generate the samples with the MH algorithm was  approximately 1 minute.

            \subsubsection{Conductivity values parametrization}
            \label{subsubsection-results-conductivity-values-parametrization}

                In this section we will discuss the conductivity-value-based parametrization with both uniform and normal priors. For each one of these cases, we tested different initial guesses for the conductivity values  $\kappa_n$.
                We did this in order to obtain insights about the reliability and robustness of our method.
                Since the conductivities are positive, these different initial guesses are also positive.
                For brevity, we show only one set of Markov chains for each prior, where we set the initial guesses for each $\kappa_n$ equal to 1.

\begin{figure}
                    \centering
                    \includegraphics[width=0.45\linewidth]{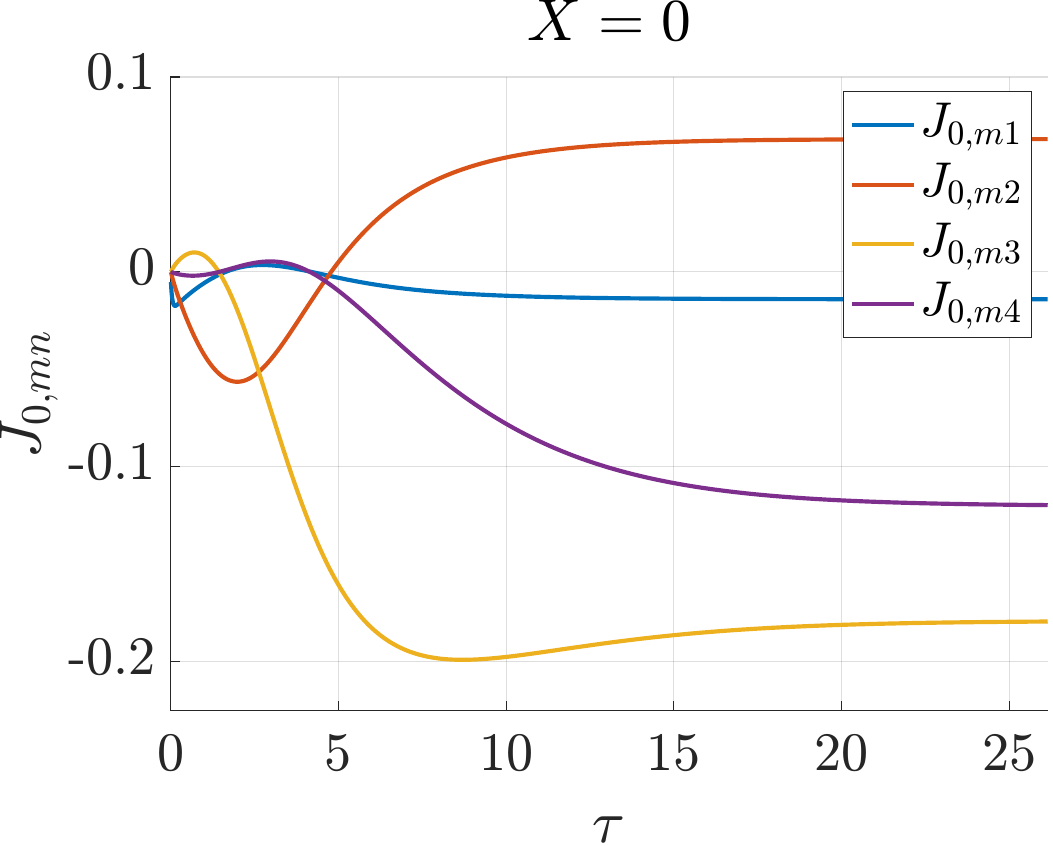}
                    \hfil
                    \includegraphics[width=0.45\linewidth]{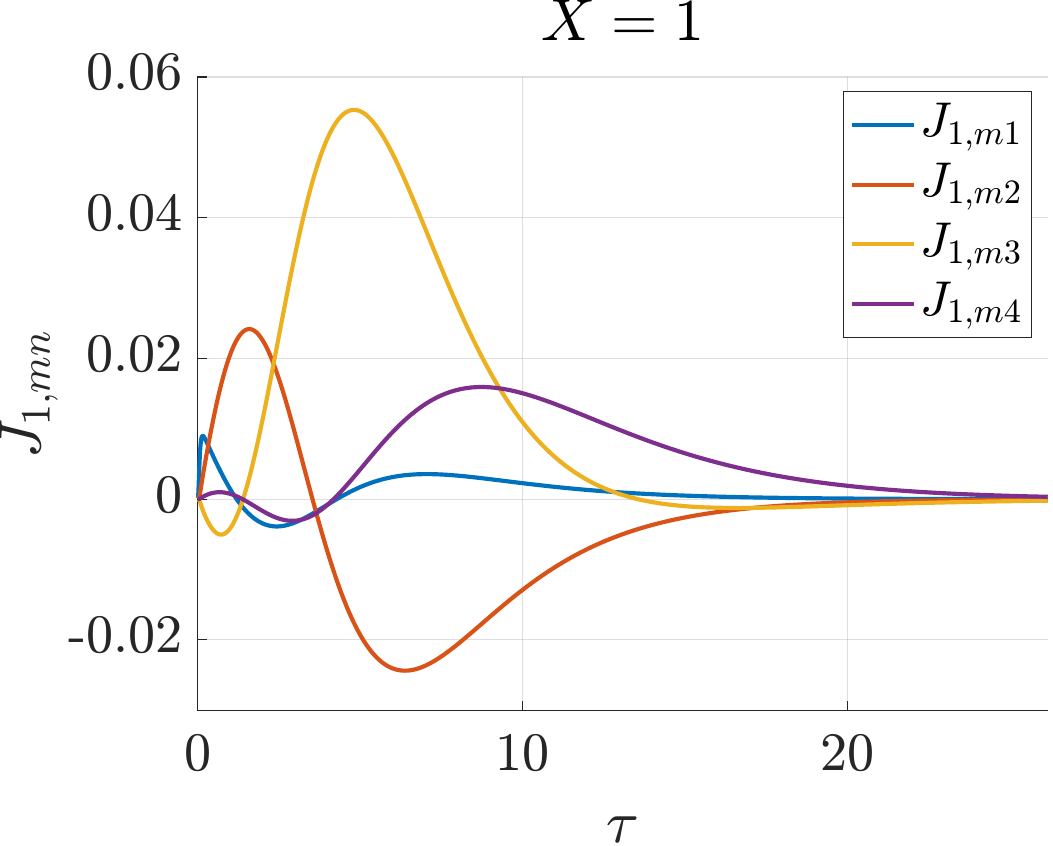}
                    \caption{On the left, sensitivity coefficients $J_{0, mn}$ with respect to $\kappa_n$ evaluated at $X = 0$.
                    On the right, sensitivity coefficients $J_{1, mn}$ with respect to $\kappa_n$ evaluated at $X = 1$.}
                    \label{figure-sensitivity-coefficients}
                \end{figure}

                \autoref{figure-sensitivity-coefficients} illustrates the sensitivity coefficients $J_{0,mn}$ and $J_{1,mn}$, with respect to the conductivities $\kappa_n$.
                These sensitivity coefficients are respectively evaluated at the left and right edges of the slab, and the reference values of $\kappa_n$ are set to the true ones.
                This results in $\JTJ = 11.56$, which indicates that all the conductivities can be simultaneously estimated.
                We notice that all the sensitivity coefficients remain constant after the system reaches the steady state condition.
                The sensitivity coefficients at $X = 0$ have a larger magnitude when compared to those at $X = 1$, which indicates that the data obtained at the left edge is more informative and improves the accuracy of the estimation of $\kappa_n$.
                The sensitivity coefficients $J_{0,m1}$ and $J_{1,m1}$ have the smallest magnitudes, which indicates that the estimation of $\kappa_1$ can involve larger uncertainties when compared to the estimation of $\kappa_2$, $\kappa_3$ and $\kappa_4$.
                This occurs because essentially the same value of the temperature would be obtained for large changes in $\kappa_1$.
                Next to that, we also see that all the sensitivity coefficients evaluated at $X = 1$ tend to zero once the system reaches the steady-state condition.
                This happens because, once the system reaches such condition, the right edge of the slab ($X = 1$) becomes governed by the heat transfer coefficient and the air temperature, and therefore changes in the conductivity do not result in relevant changes in the local temperature.

                In order to obtain insights about the sensitivity analysis when the reference values for $\kappa_n$ are significantly different from the true ones, preliminary simulations were also performed by setting these reference values to 1.
                We selected 1 because it has the same order of magnitude as the true values of the conductivities.
                The obtained sensitivity coefficients, which are omitted here for brevity, follow the same trends as those in \autoref{figure-sensitivity-coefficients}, but with larger magnitudes.
                Additionally, we obtained for this case $\JTJ = \num{2.81e7}$, which again indicates that all conductivities can be simultaneously estimated.
                Hence, the sensitivity analysis shows us that, even if we don't know the true values of the parameters we want to estimate, it is possible to use reference values and obtain insights about how to conduct the inference procedure, such as defining the optimal sensor location and understanding which parameters can be simultaneously estimated.

\begin{figure}
                    \centering
                    \includegraphics[width=\linewidth]{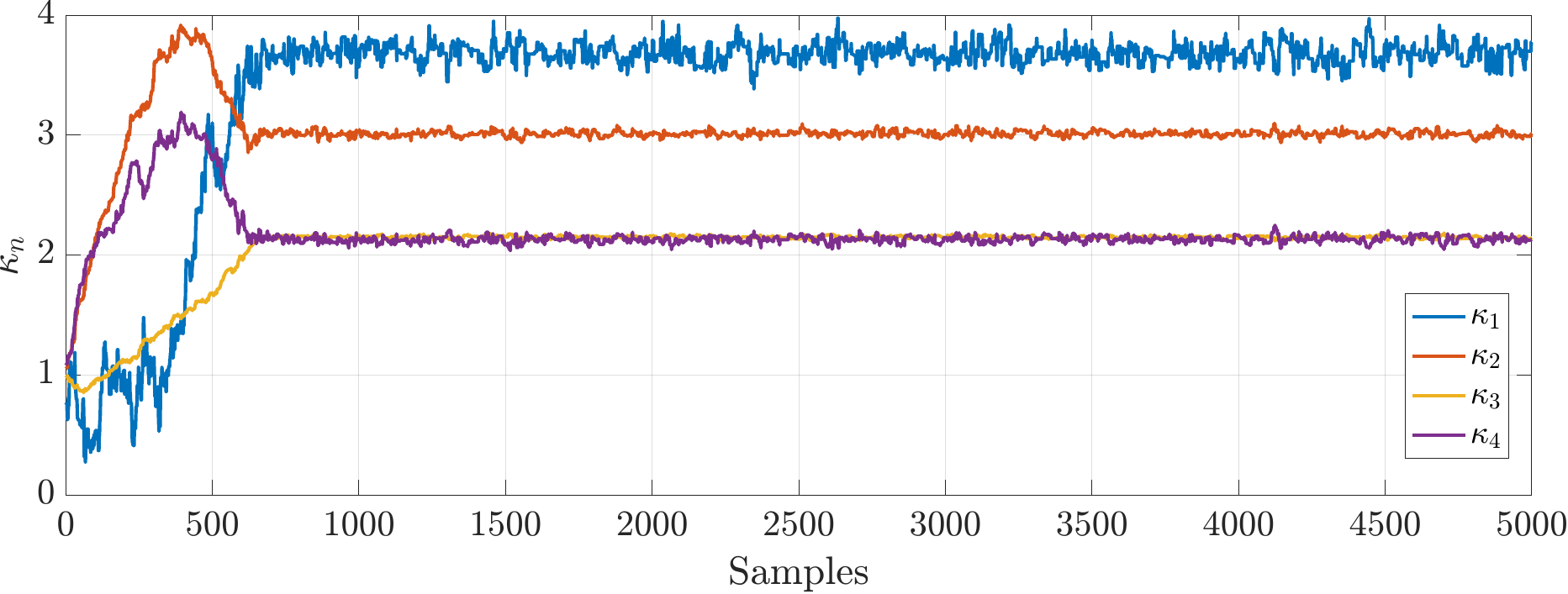}
                    \caption{Markov chains of the conductivities $\kappa_n$ obtained with the uniform prior and the third-degree polynomial model parametrized with the conductivity values.}
                    \label{figure-markov-chains-k}
                \end{figure}

\begin{figure}
                    \centering
                    \includegraphics[width=0.45\linewidth]{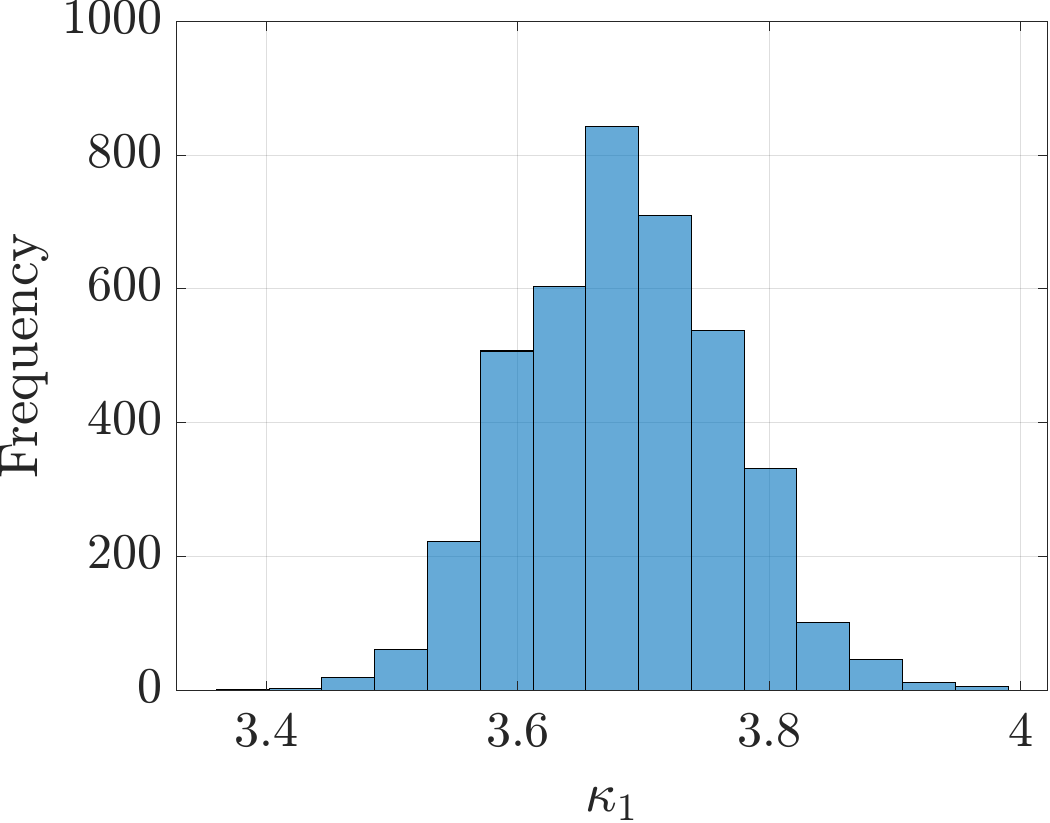}
                    \hfil
                    \includegraphics[width=0.45\linewidth]{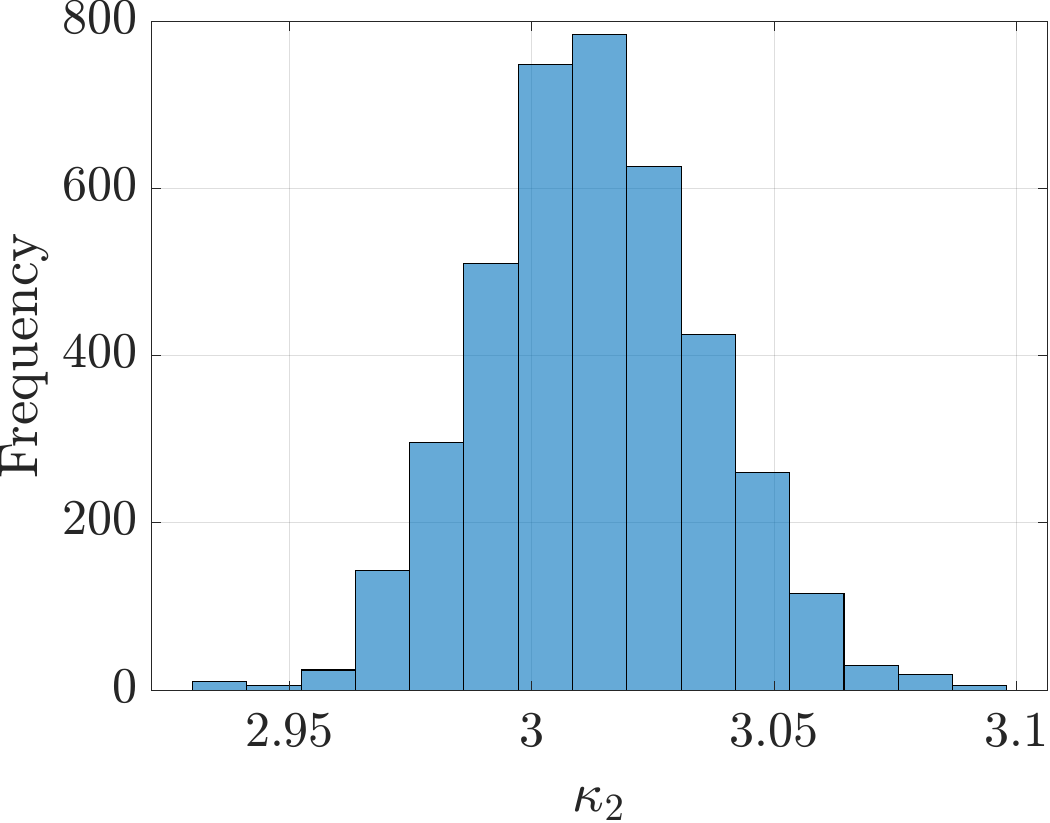}
                    \includegraphics[width=0.45\linewidth]{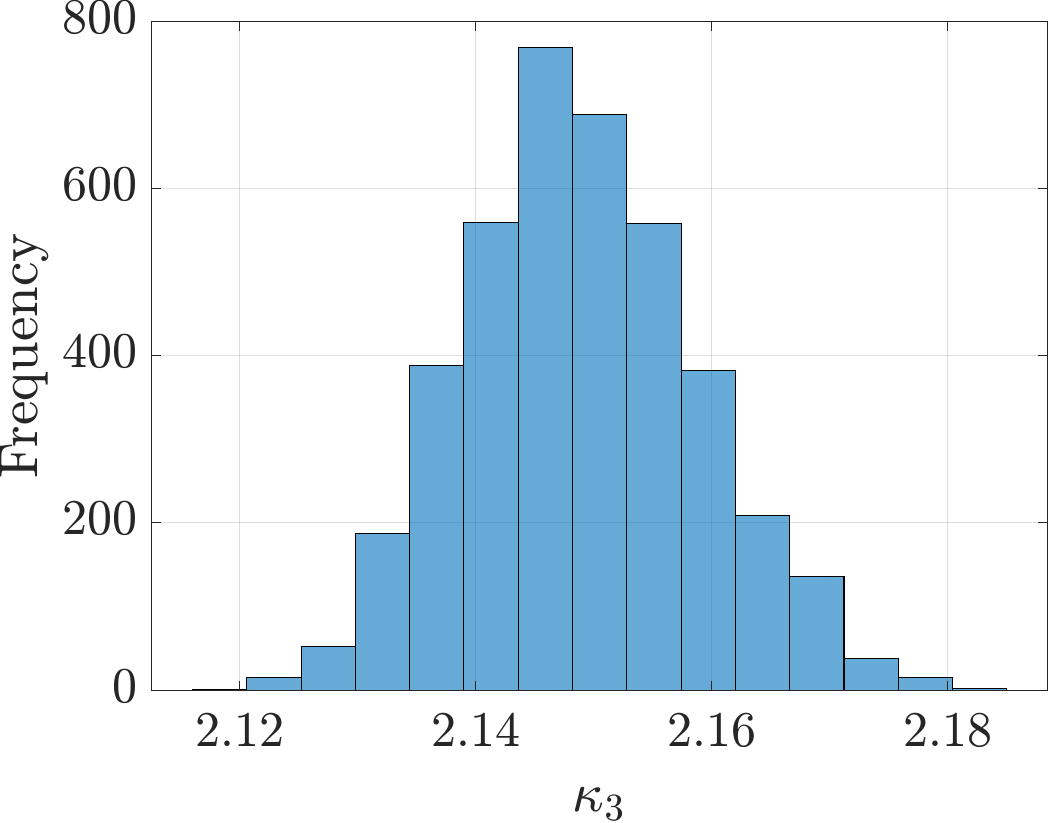}
                    \hfil
                    \includegraphics[width=0.45\linewidth]{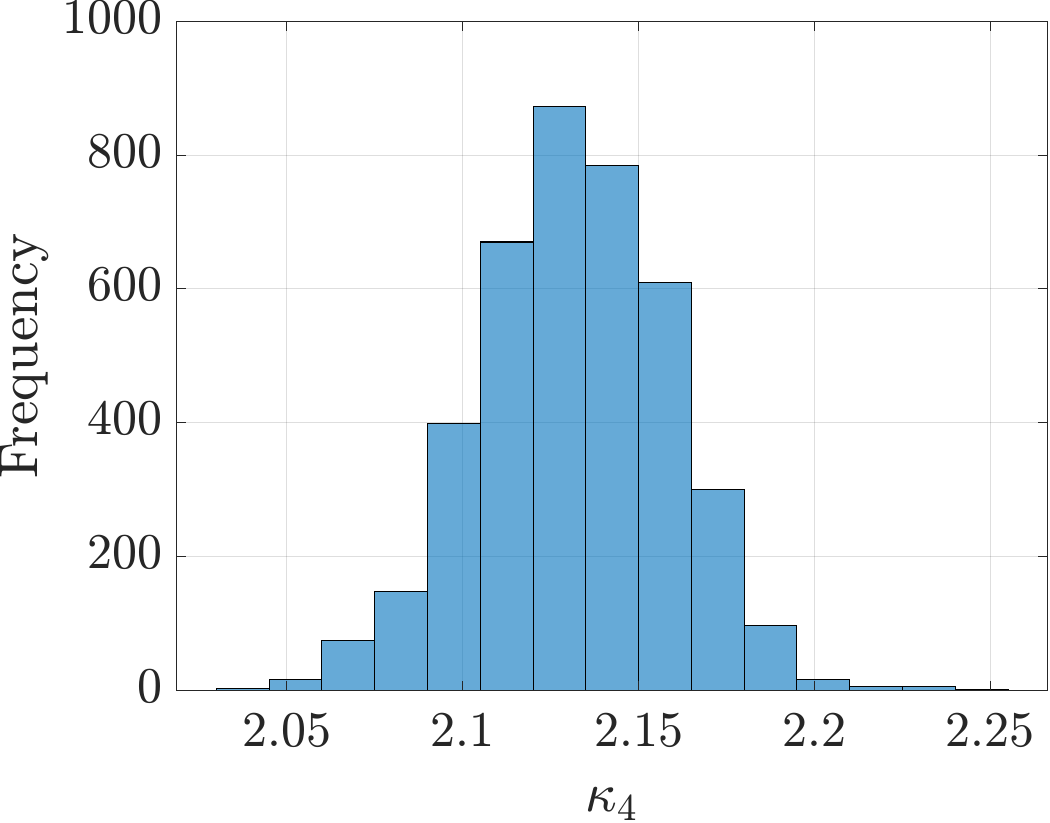}
                    \caption{Histograms of the limit distributions regarding the conductivities $\kappa_n$ obtained with the uniform prior and the third-degree polynomial model parametrized with the conductivity values.}
                    \label{figure-histograms-k}
                \end{figure}

                Figure \ref{figure-markov-chains-k} illustrates the Markov chains with respect to the conductivities $\kappa_n$ obtained with the uniform improper prior.
                It is clear that the selection of the first \num{1000} samples is appropriate to define the burn-in period, since each chain requires between \num{500} and \num{1000} samples to reach equilibrium.
                The remaining samples satisfy the convergence criterion and are used to represent the limit distribution and therefore the posterior.
                The histograms of the samples of the limit distributions are illustrated in \autoref{figure-histograms-k}.
                The relevant statistics of these samples, as well as the true values of $\kappa_n$, are listed in \autoref{table-statistics-k}.
                We can see from this table that standard deviations are small when compared to the respective mean values, thus resulting in small relative standard deviations.
                This means that the conductivities $\kappa_n$ were estimated with small uncertainties.
                \autoref{table-statistics-k} also shows the relative error, which is defined as $|(\kappa_m - \kappa_t) / \kappa_t|$, where $\kappa_m$ and $\kappa_t$ represent respectively the mean and the true value.
                We can notice that the conductivity $\kappa_1$ has the largest relative standard deviation and relative error.
                This was already expected, since the sensitivity coefficients (\autoref{figure-sensitivity-coefficients}) for this quantity have the smaller magnitudes when compared to the sensitivity coefficients with respect to the other three conductivities.

                \begin{table}
                    \centering
                    \caption{True values of $\kappa_n$ and statistics of the limit distributions regarding these quantities obtained with the uniform prior and the third-degree polynomial model parametrized with the conductivity values.}
                    \label{table-statistics-k}
                    \begin{tabular}{lcccc}
                        \toprule
                        & $\kappa_1$ & $\kappa_2$ & $\kappa_3$ & $\kappa_4$
                        \\
                        \midrule
                        True values & 3.8928 & 2.9689 & 2.1350 & 2.1146
                        \\
                        Mean & 3.6851 & 3.0079 & 2.1474 & 2.1323
                        \\
                        Standard deviation & 0.0851 & 0.0239 & 0.0100 & 0.0279
                        \\
                        Relative standard deviation (\%) & 2.3106 & 0.7960 & 0.4675 & 1.3104
                        \\
                        Relative error (\%) & 5.3355 & 1.3133 & 0.5825 & 0.8351
                        \\
                        \bottomrule
                    \end{tabular}
                \end{table}

\begin{figure}
                    \centering
                    \includegraphics[width=\linewidth]{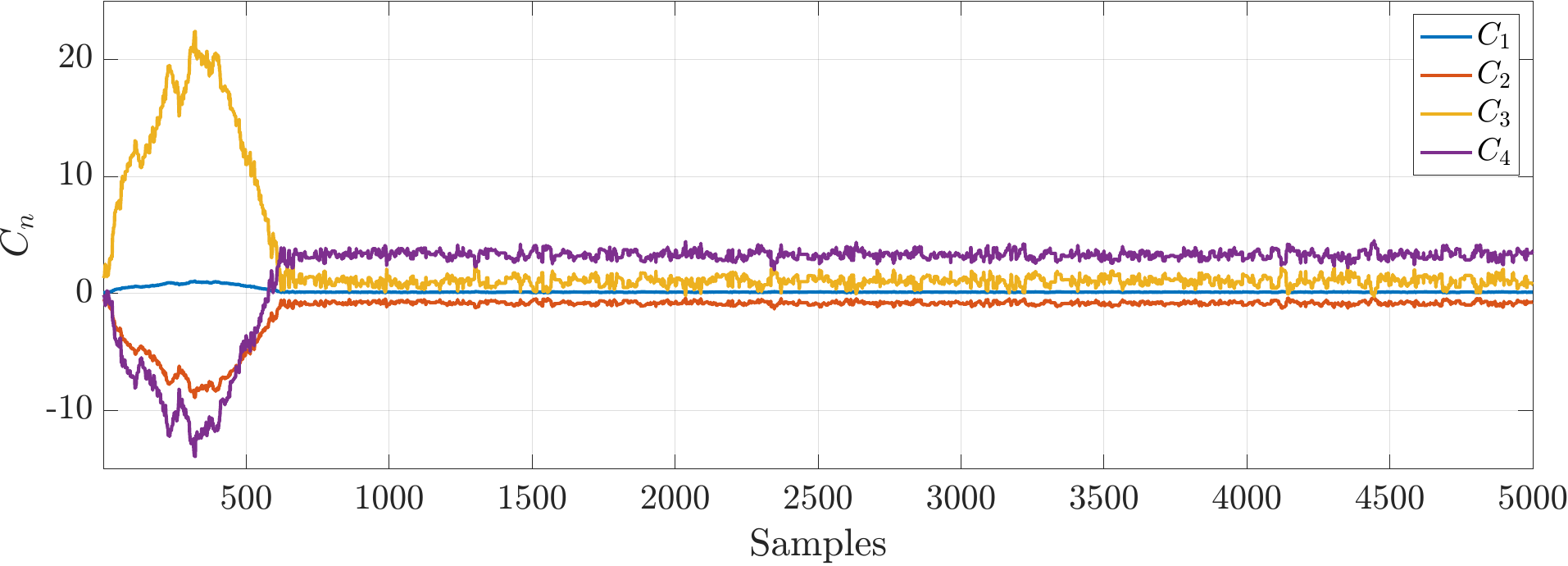}
                    \caption{Markov chains of the coefficients $C_n$ obtained with the uniform prior and the third-degree polynomial model parametrized with the conductivity values.}
                    \label{figure-markov-chains-C}
                \end{figure}

\begin{figure}
                    \centering
                    \includegraphics[width=0.45\linewidth]{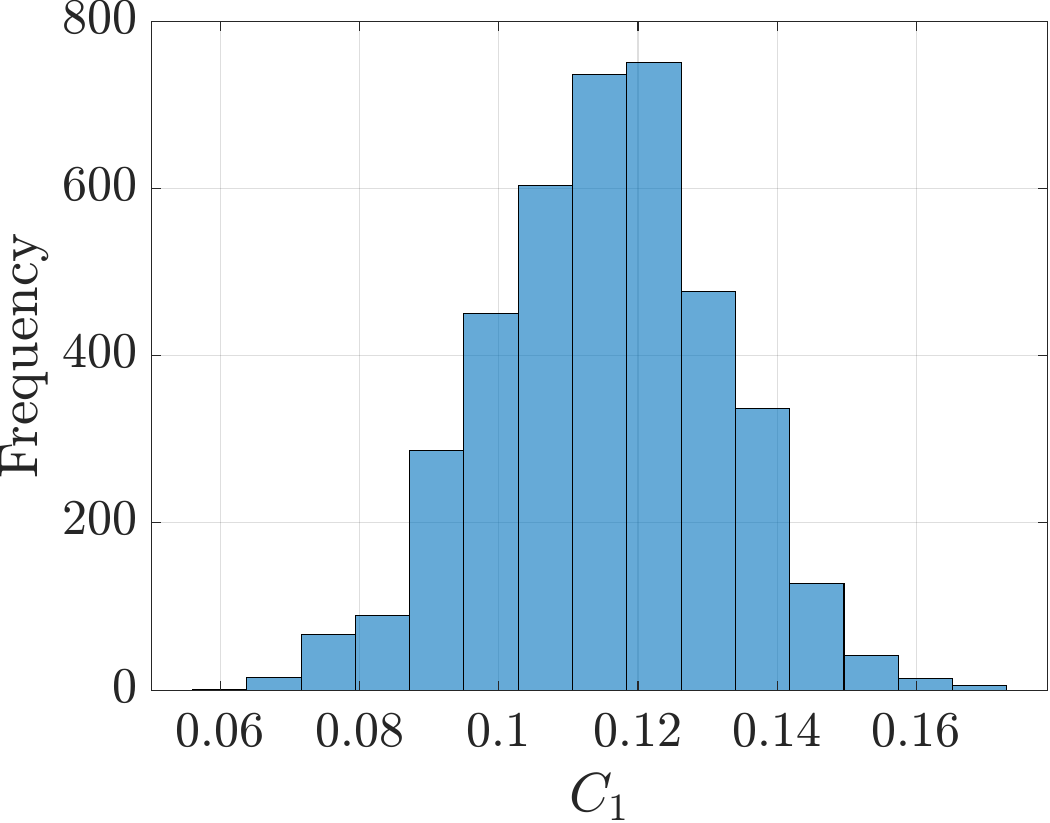}
                    \hfil
                    \includegraphics[width=0.45\linewidth]{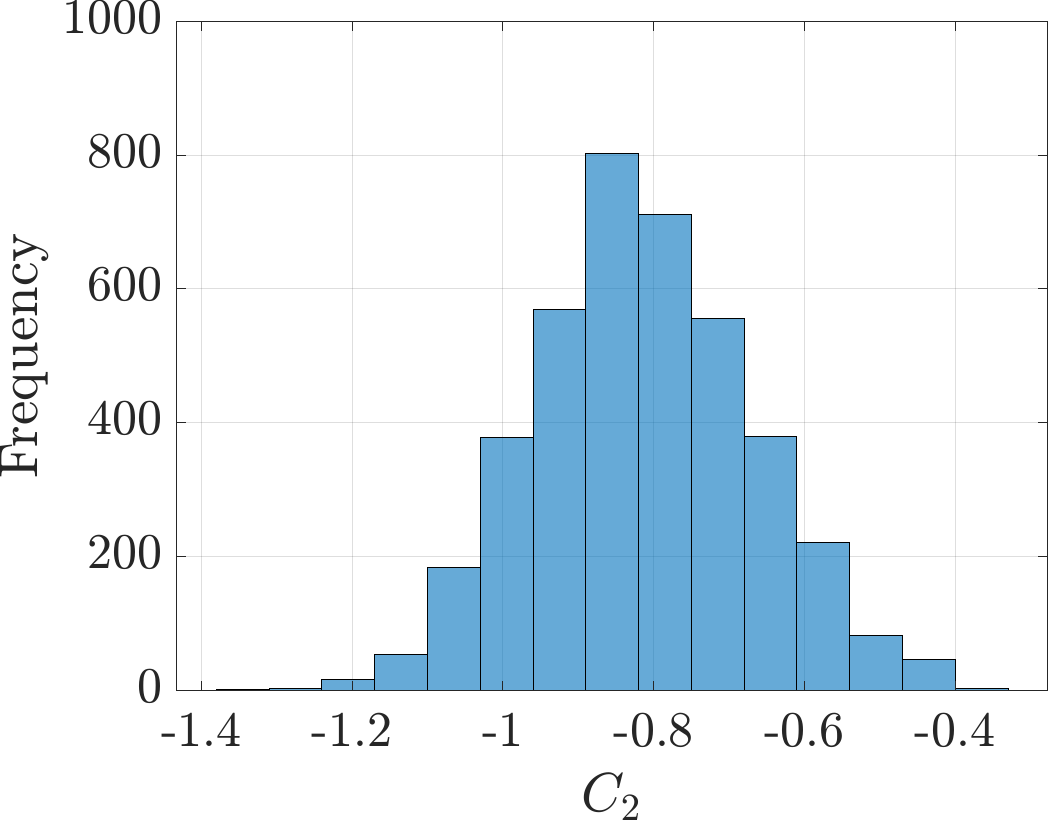}
                    \includegraphics[width=0.45\linewidth]{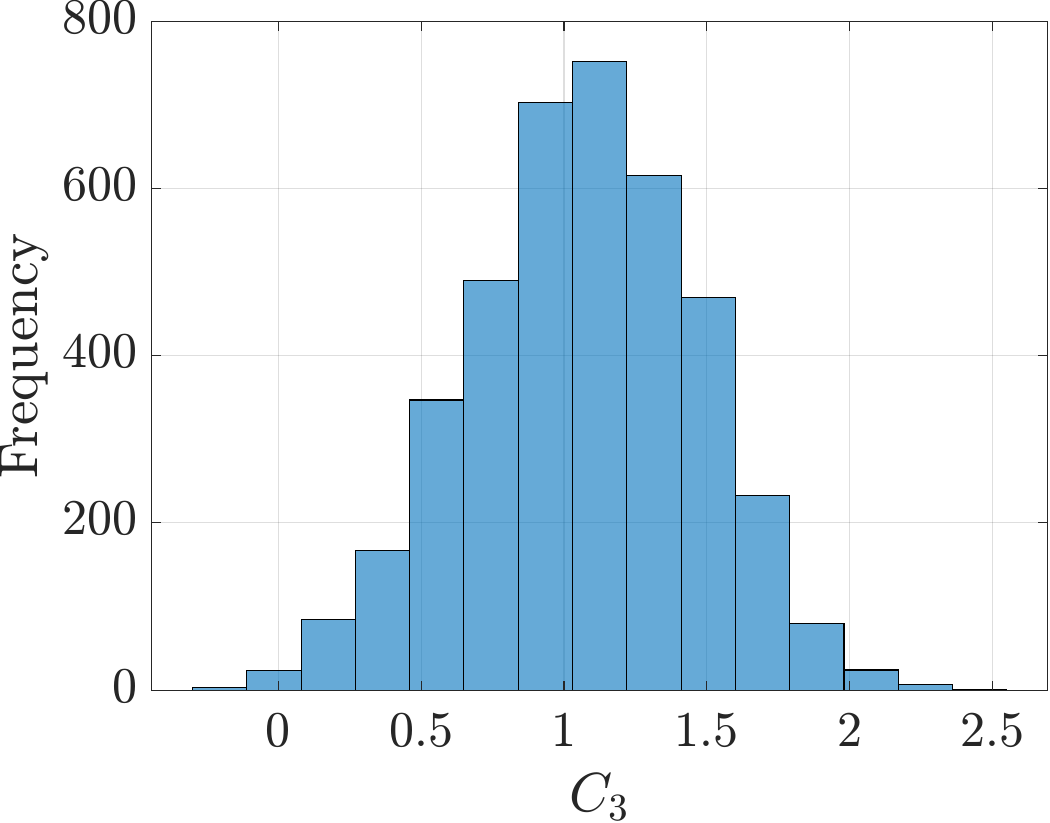}
                    \hfil
                    \includegraphics[width=0.45\linewidth]{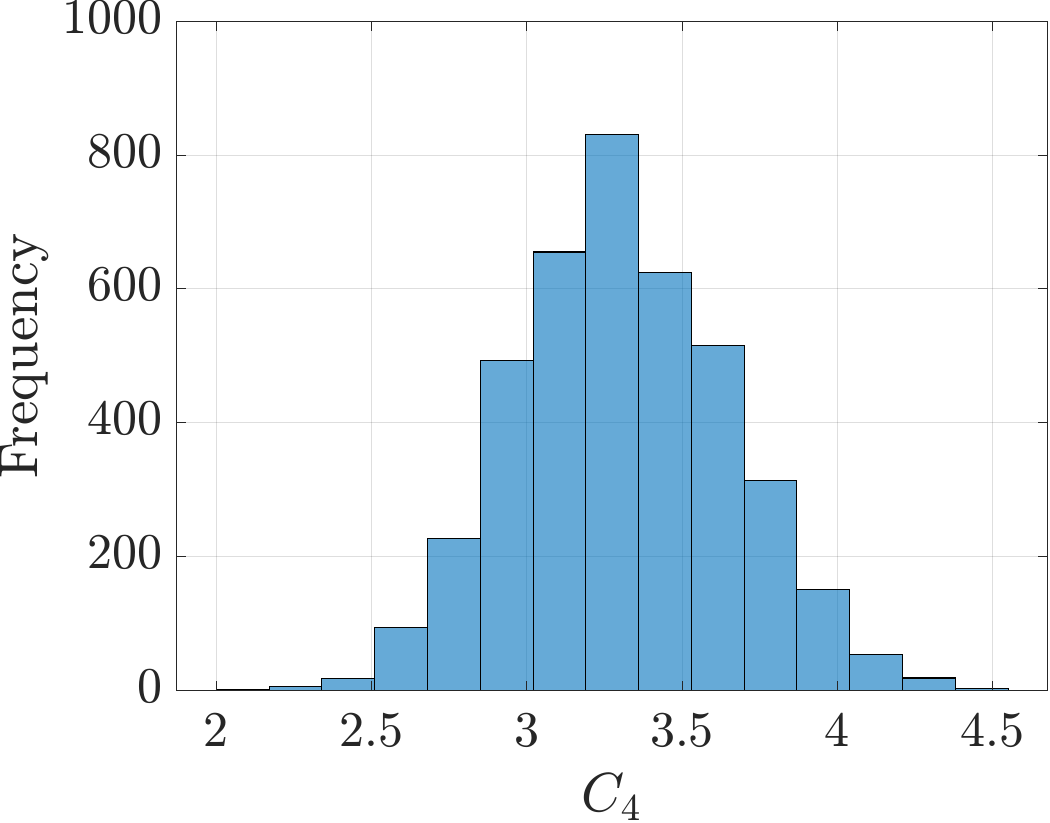}
                    \caption{Histograms of the limit distributions regarding the coefficients $C_n$ obtained with the uniform prior and the third-degree polynomial model parametrized with the conductivity values.}
                    \label{figure-histograms-C}
                \end{figure}

                Each set of the conductivities $\kappa_n$ obtained from the Markov chains shown in \autoref{figure-markov-chains-k} is associated to a set of coefficients $C_n$.
                Hence, we can also obtain Markov chains regarding these coefficients and perform the same analysis regarding their statistics.
                \autoref{figure-markov-chains-C} illustrates the Markov chains with respect to the coefficients $C_n$.
                We notice again that all chains reach a limit distribution.
                The histograms of the samples from the limit distribution are shown in \autoref{figure-histograms-C}.
                The statistics of these samples are omitted here for brevity.

\begin{figure}
                    \centering
                    \begin{subfigure}[t]{0.45\linewidth}
                        \vskip 0pt
                        \includegraphics[width=\linewidth]{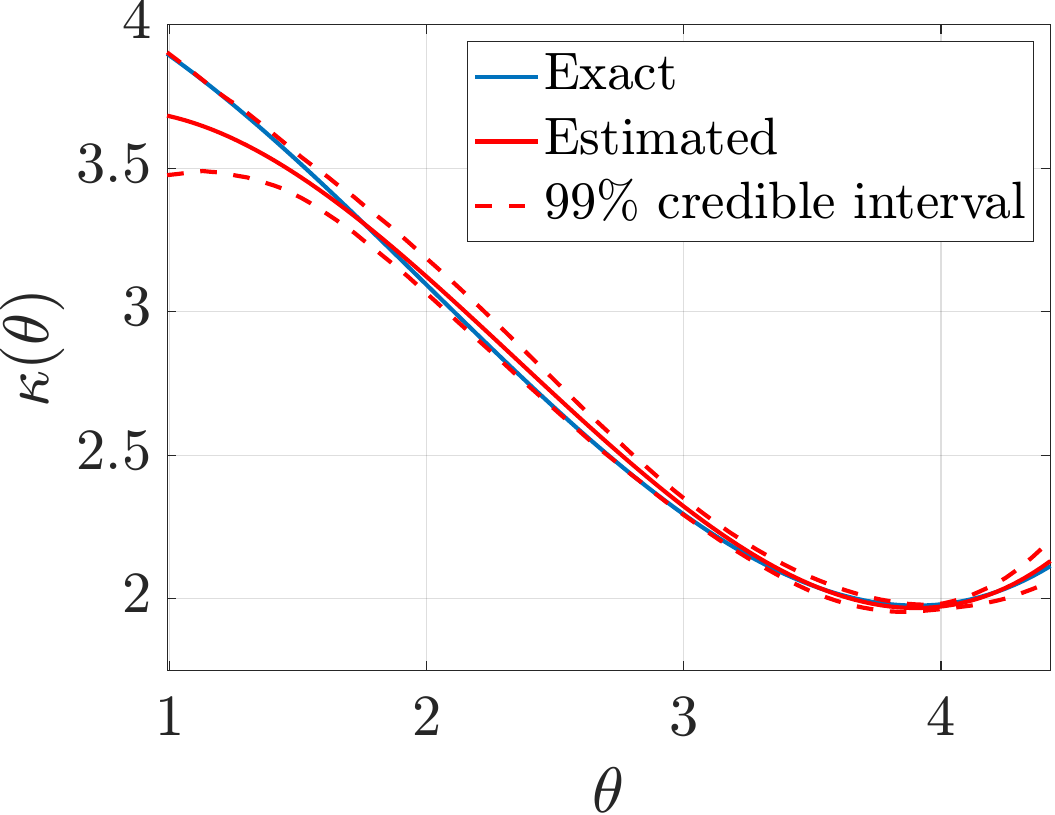}
                        \caption{Conductivity values parametrization and uniform prior.}
                        \label{figure-estimation-and-uq-values-uniform}
                    \end{subfigure}
                    \hfil
                    \begin{subfigure}[t]{0.45\linewidth}
                        \vskip 0pt
                        \includegraphics[width=\linewidth]{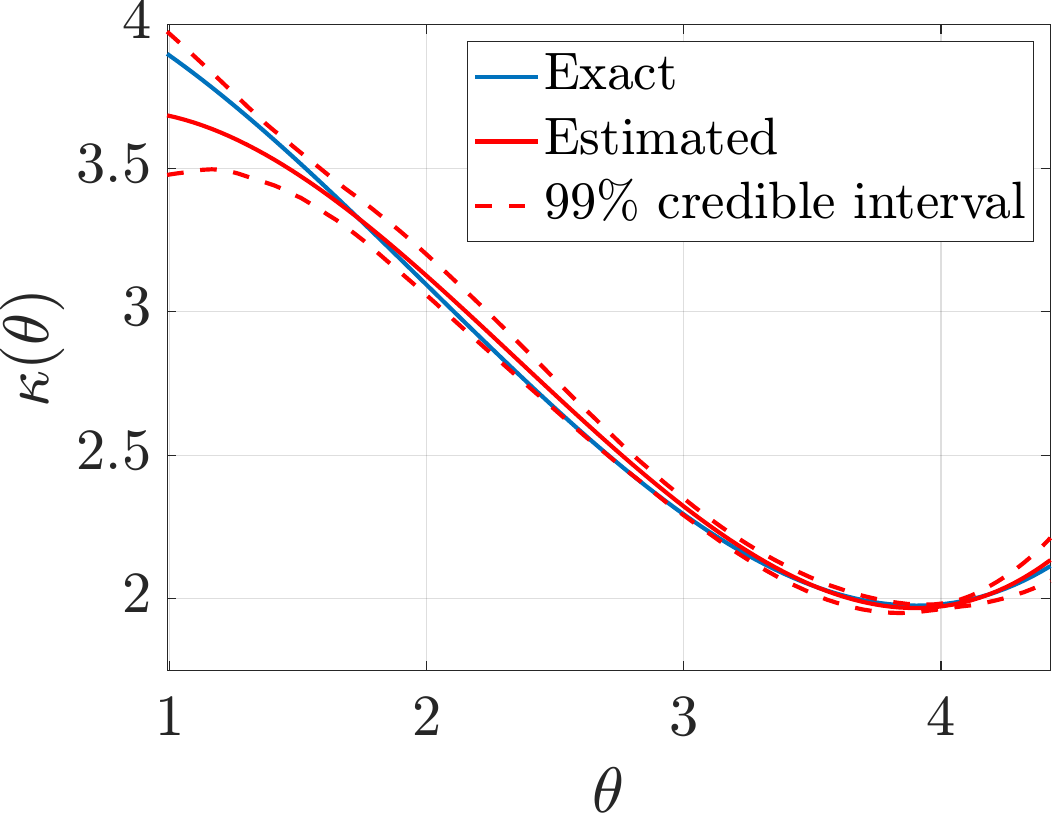}
                        \caption{Coefficients parametrization and uniform prior.}
                        \label{figure-estimation-and-uq-coeff-uniform}
                    \end{subfigure}
                    \par\medskip
                    \begin{subfigure}[t]{0.45\linewidth}
                        \vskip 0pt
                        \includegraphics[width=\linewidth]{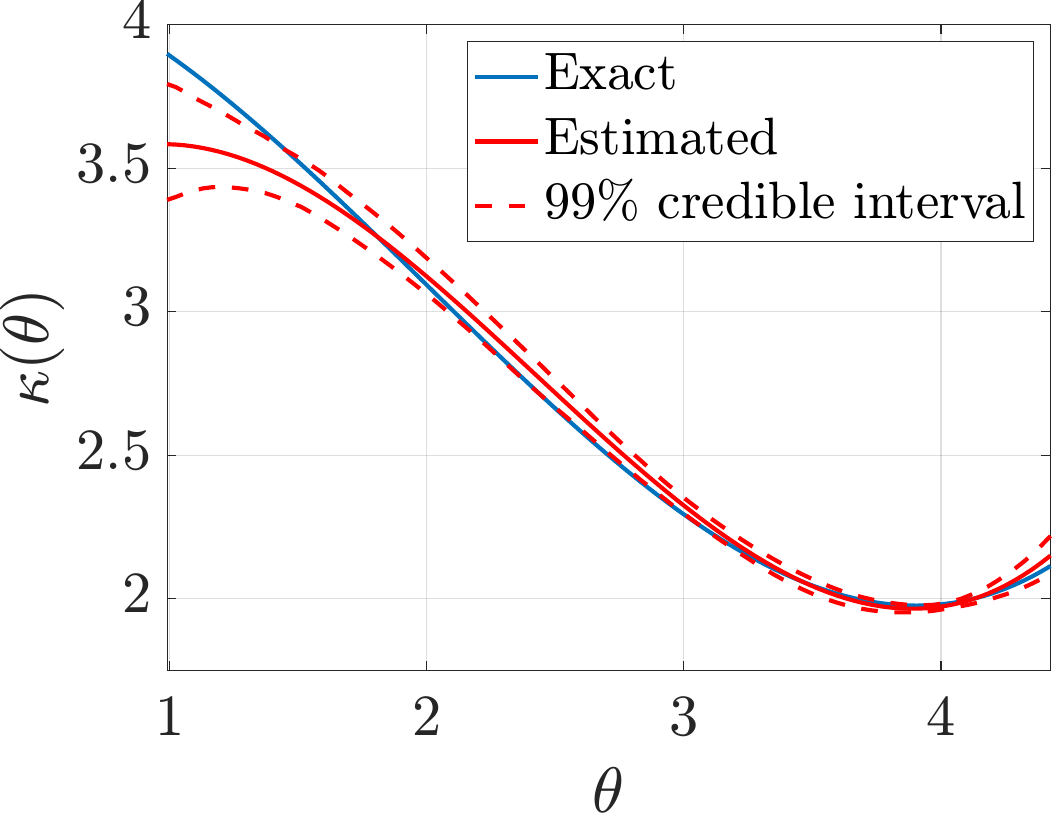}
                        \caption{Conductivity values parametrization and normal prior with relative standard deviation of 10\%.}
                        \label{figure-estimation-and-uq-values-normal10}
                    \end{subfigure}
                    \hfil
                    \begin{subfigure}[t]{0.45\linewidth}
                        \vskip 0pt
                        \includegraphics[width=\linewidth]{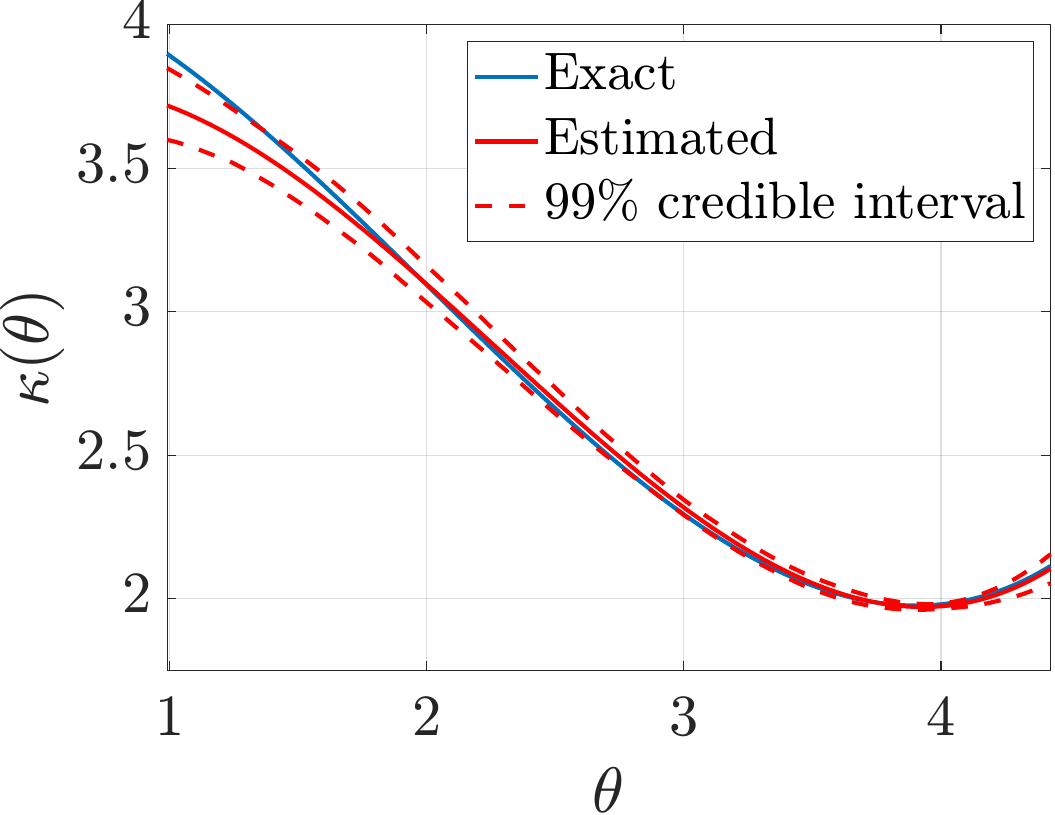}
                        \caption{Coefficients parametrization and normal prior with relative standard deviation of 10\%.}
                        \label{figure-estimation-and-uq-coeff-normal10}
                    \end{subfigure}
                    \par\medskip
                    \begin{subfigure}[t]{0.45\linewidth}
                        \vskip 0pt
                        \includegraphics[width=\linewidth]{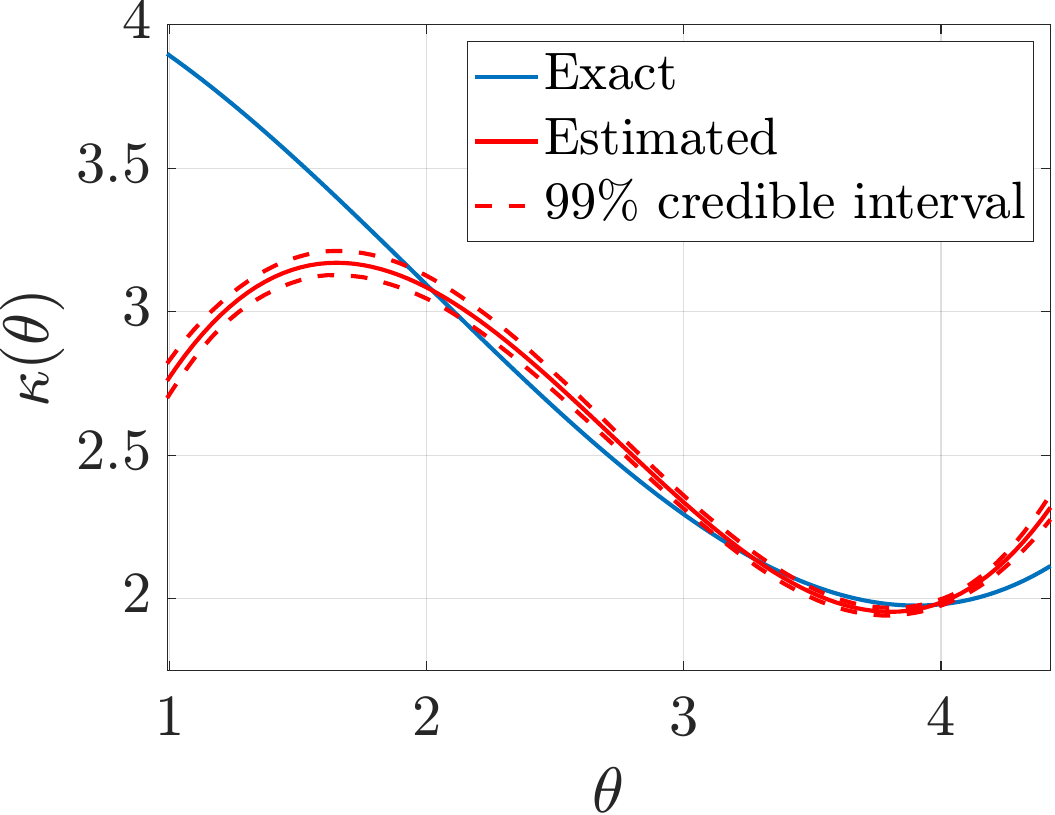}
                        \caption{Conductivity values parametrization and normal prior with relative standard deviation of 1\%.}
                        \label{figure-estimation-and-uq-values-normal1}
                    \end{subfigure}
                    \hfil
                    \begin{subfigure}[t]{0.45\linewidth}
                        \vskip 0pt
                        \includegraphics[width=\linewidth]{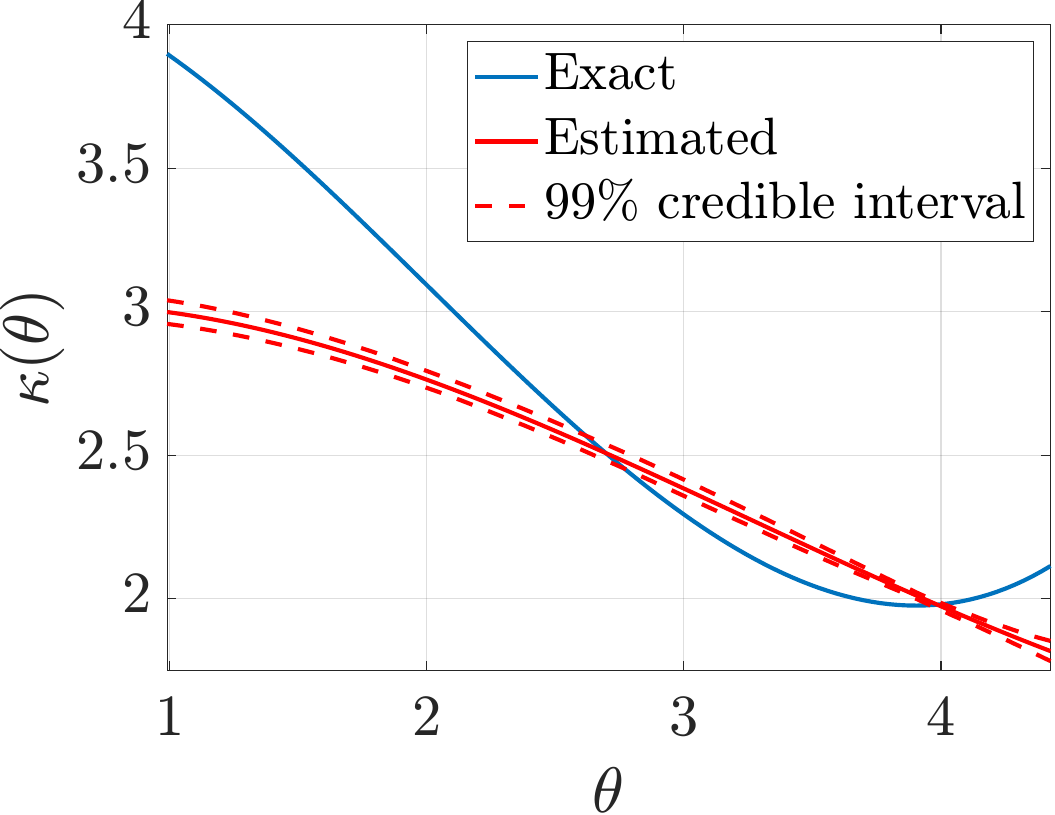}
                        \caption{Coefficients parametrization and normal prior with relative standard deviation of 1\%.}
                        \label{figure-estimation-and-uq-coeff-normal1}
                    \end{subfigure}
                \caption{Comparisons between the exact and estimated values of $\kappa(\theta)$ obtained with the third-degree polynomial model, including credible intervals.
                }
                \end{figure}

                Once having the values of the coefficients $C_n$ for each state, we can model the function $\kappa(\theta)$.
                Since each state assumes specific values for $\kappa(\theta)$, we can not only obtain the expected values of this function, but also quantify its uncertainties.
                \autoref{figure-estimation-and-uq-values-uniform} shows a comparison between the exact and expected values of $\kappa(\theta)$ obtained with the third-degree polynomial model and the uniform prior.
                This figure also quantifies the uncertainty through the 99\% credible interval.
                We observe a good agreement between both exact and expected values.
                Besides, the exact values are located within the 99\% credible interval.
                We also observe larger uncertainties for values of $\theta$ close to 1. These larger uncertainties can be explained by the sensitivity coefficients (\autoref{figure-sensitivity-coefficients}).
                One should recall here that $\theta = 1$ represents the initial condition of our system.
                At the left edge ($X = 0$) we have the sensitivity coefficients with the largest magnitudes, which means that the temperature values obtained at this edge are more informative for the solution than those obtained at the right edge.
                Additionally, we can see that, also for $X = 0$, the sensitivity coefficients have small magnitudes for small values of the time $\tau$, where the temperature is still close to the initial condition.
                Since the sensitivity coefficients are small when $\theta$ is close to 1, the uncertainties regarding the estimates of $\kappa_n$ and $C_n$ increase and therefore we obtain larger uncertainties for the function $\kappa(\theta)$.

                We now focus on the results obtained with normal priors.
                We assume that these priors have the same mean vector $\bm{\mu} = [\mu_1, \mu_2, \mu_3, \mu_4]^T$, where $\mu_n$ denotes the mean corresponding to $\kappa_n$.
                For simplicity, we assume that all $\mu_n$ have the same value, namely $\mu$.
                We set this value as the mean of the function used to generate the simulated measurements within the temperature range of interest, that is, $\mu = (\theta_{\text{max}} - \theta_{\text{min}})^{-1} \int_{\theta_{\text{min}}}^{\theta_{\text{max}}} \sum_{n=1}^4 C_n \theta^{4-n} \, d\theta$, where the quantities $C_n$ represent the true coefficients of the third-degree polynomial model shown in \autoref{subsection-definition-of-simulated-temperature-measurements}.
                This results in $\mu \approx 2.66$.
                Additionally, for the definition of the covariance matrix $\bm{V}$, we select two different relative standard deviations: 10\% and 1\%.
                
                The value of $\mu$ selected above is just a reference and should be adjusted according to the problem under analysis.
                Besides, the distributions for each conductivity $\kappa_n$ could also assume non-identical means and standard deviations.
                This is convenient for situations where we have prior knowledge about the values of the conductivity at specific temperatures.
                For example, if we have a strong prior knowledge about the value of $\kappa_1$, which represents the conductivity when the temperature is equal to the initial condition, we can define a specific mean $\mu_1$ and a small standard deviation $\sigma_1$.
                If there is a lack of information about how the conductivity $\kappa_4$ behaves at the maximum measured temperature, we can define another mean value $\mu_4$ and a large standard deviation $\sigma_4$.

                Figures with the Markov chains and the histograms obtained for a normal prior are very similar to those obtained using the uniform prior and are therefore omitted here for brevity.
                \autoref{figure-estimation-and-uq-values-normal10} illustrates the comparison between the exact and expected values of $\kappa(\theta)$ obtained with the third-degree polynomial model and the normal prior with relative standard deviation of 10\%.
                This figure also shows the 99\% credible interval.
                We can notice that, in contrast to the results obtained with the uniform prior (\autoref{figure-estimation-and-uq-values-uniform}), the exact values of $\kappa$ for values of $\theta$ close to 1 are not located within the credible interval.
                This happens because the mean value used in the prior is smaller than the exact values of $\kappa(\theta)$.
                Additionally, the relative standard deviation of 10\% is not sufficiently large to provide results in which the exact values are located within the credible intervals.

                The comparison between the exact and expected values of $\kappa(\theta)$ for a relative standard deviation of 1\% is illustrated in \autoref{figure-estimation-and-uq-values-normal1}.
                We notice a smaller credible interval when compared to the previous results.
                This occurs because now we consider a relative standard deviation that is 10 times smaller.
                We also notice a strong disagreement between the exact and expected values of $\kappa(\theta)$, specially for $\theta$ located between 1 and 2.
                This occurs because the mean value is different from the exact ones, at the same time that the standard deviation is very small, thus resulting in a strongly biased prior.
                These results also show that informative priors must be carefully selected.
                Although these priors provide small credible intervals, they can lead to wrong interpretations in case the mean values are not properly selected.

            \subsubsection{Coefficients parametrization}
            \label{subsubsection-results-coeficient-parametrization}

                We now discuss the results obtained with the coefficients parametrization. We omit the Markov chains and histograms of the samples of the limit distributions, as these results do not provide additional insights compared to the results for the parametrization using conductivity values as discussed above. Instead, we focus our analysis on understanding how the parametrization alters the solution of the inverse problem.
                
                \autoref{figure-estimation-and-uq-coeff-uniform} shows the comparison between the exact and expected values of $\kappa(\theta)$ obtained with the coefficients parametrization and the uniform prior.
                We observe that these results closely resemble those obtained with the conductivity values parametrization (\autoref{figure-estimation-and-uq-values-uniform}).
                This resemblance is explained by the considered non-informative prior, for which the solution of the inverse problem becomes dominated by the likelihood. As a consequence, the influence of the prior, and specifically the way in which it is parametrized, has a limited effect on the inference.

                We also study the results obtained with informative normal priors.
                In contrast to the parametrization by means of conductivity values, selecting priors directly for the coefficients is more intricate.
                We here select the priors by choosing a mean vector that favors values of $\kappa(\theta)$ around the same average value as used in the conductivity values parametrization, i.e., $\mu \approx 2.66$.
                We achieve this by setting the mean of $C_4$ to $\mu$ and those of $C_1$, $C_2$ and $C_3$ to 0.
                The mean vector is then defined as $\bm{\mu} = (0, 0, 0, \mu)^T$. Furthermore, we select the same standard deviation for all coefficients, which is equal to 10\% or 1\% of $\mu$.
                
                Figures \ref{figure-estimation-and-uq-coeff-normal10} and \ref{figure-estimation-and-uq-coeff-normal1} show the comparison between the exact and estimated values of $\kappa(\theta)$ obtained with the coefficients parametrization and the normal prior with standard deviations equal to 10\% and 1\% of $\mu$, respectively.
                The results obtained with a standard deviation equal to 10\% of $\mu$ do not substantially differ from those obtained with the conductivity values parametrization (\autoref{figure-estimation-and-uq-values-normal10}).
                Similar to the uniform prior case discussed above, this is a consequence of the prior being sufficiently uninformative, such that the solution of the inverse problem is dominated by the likelihood.
                By setting the standard deviation to 1\% of $\mu$ (\autoref{figure-estimation-and-uq-coeff-normal1}), a strong disagreement between the exact and estimated values of $\kappa(\theta)$ is observed.
                This occurs because the prior has a mean different from the exact values and a small standard deviation, which significantly biases the posterior estimates away from the ground truth.
                Furthermore, these results differ from those obtained with the conductivity values parametrization (\autoref{figure-estimation-and-uq-values-normal1}), for which a good resemblance is still attained in the range of the conductivity values where the informed prior is reasonably accurate. In contrast, in the case of the coefficients parametrization, the effect of the prior bias spreads out over the complete temperature range.
                When using the coefficients parametrization, the solution to the inverse problem is observed to be more sensitive to the quality of informative priors compared to the parametrization based on the conductivity values.

        \subsection{Piecewise linear functions model}

            We discretize the function $\kappa(\theta)$ with a total of 100 points.
            This means we want to estimate a total of $N = 100$ values of the conductivity.
            We select this number with the argument that it provides a good balance between the quality of the approximation and computation cost.
            We show below the results obtained when the initial guesses for $\kappa_n$ are set to 1.
            In a similar way to what is shown in the previous section, we also tested different initial guesses, and they resulted in the same limit distribution.
            We omit the Markov chains here for brevity.
            For all the cases discussed below, we consider \num{1000000} and \num{50000} samples for the adaptive MCMC and MH algorithms, respectively.
            The computational times required to generate these samples were respectively 5 hours and 15 minutes.
            
            In order to use the GMRF prior, we need to specify the values of the mean $\bar{\bm{Q}}$ and the variance $\gamma^2$.
            With the intention of understanding their effects on the solution of the inverse problem, we decide to select three different values for each one of these quantities.
            First we keep $\bar{\bm{Q}}$ constant and study the effects of three values of $\gamma^2$.
            Next, we keep $\gamma^2$ constant and vary $\bar{\bm{Q}}$.
            
            In order to obtain our first reference for $\gamma^2$, we assume that $\bar{\bm{P}}$ contains the exact values of the conductivity.
            This means that $\bar{\bm{Q}}$ contains the exact values of the difference of two consecutive elements in $\bar{\bm{P}}$.
            We then set $\gamma^2$ as the variance of $\bar{\bm{Q}}$.
            This results in $\gamma^2 = \num{2e-4}$.
            
            In practice, values of $\kappa_n$ are unknown, and therefore it is not possible to precisely estimate the parameter $\gamma^2$ with the strategy mentioned above.
            The definition of $\gamma^2$ also reflects our prior knowledge about the function $\kappa(\theta)$.
            By taking the first reference $\gamma^2 = \num{2e-4}$, we decide to select the other two values as \num{2e-3} and \num{2e-5}.
            We choose these values in order to understand how changing the order of magnitude of $\gamma^2$ affects the solution of the inverse problem.
            
            Regarding $\bar{\bm{Q}}$, our first choice is to set $\bar{\bm{Q}} = \bm{Q}_{\text{exact}}$, where $\bm{Q}_{\text{exact}}$ is the vector with the differences of two consecutive elements in $\bm{P}_{\text{exact}}$.
            This is equivalent to setting $\bar{\bm{P}} = \bm{P}_{\text{exact}}$.
            Hence, the prior is a Gaussian centered at the exact values.
            In a similar way to when we assume $\gamma^2 = \num{2e-4}$, this configuration does not represent a realistic case, since in practice we do not know the exact values of $\kappa_n$ and want to estimate them.
            Next, we define $\bar{\bm{Q}}$ as a zero vector.
            This represents a situation where we expect constant values in $\bm{P}$.
            Finally, we also show results for the case where $\bar{\bm{Q}}$ is defined by setting $\bar{\bm{P}} = - \bm{P}_{\text{exact}}$.
            By doing this, our prior models a case where changes between two consecutive values in $\bm{P}$ have the same magnitude as the real ones, but in the opposite direction.
            It is evident that this selection of $\bar{\bm{Q}}$ is far from ideal, and we are interested in understanding the effects of this poor choice for our prior on the results of the inverse problem.

\begin{figure}
                \centering
                \begin{subfigure}{0.45\linewidth}
                    \includegraphics[width=\linewidth]{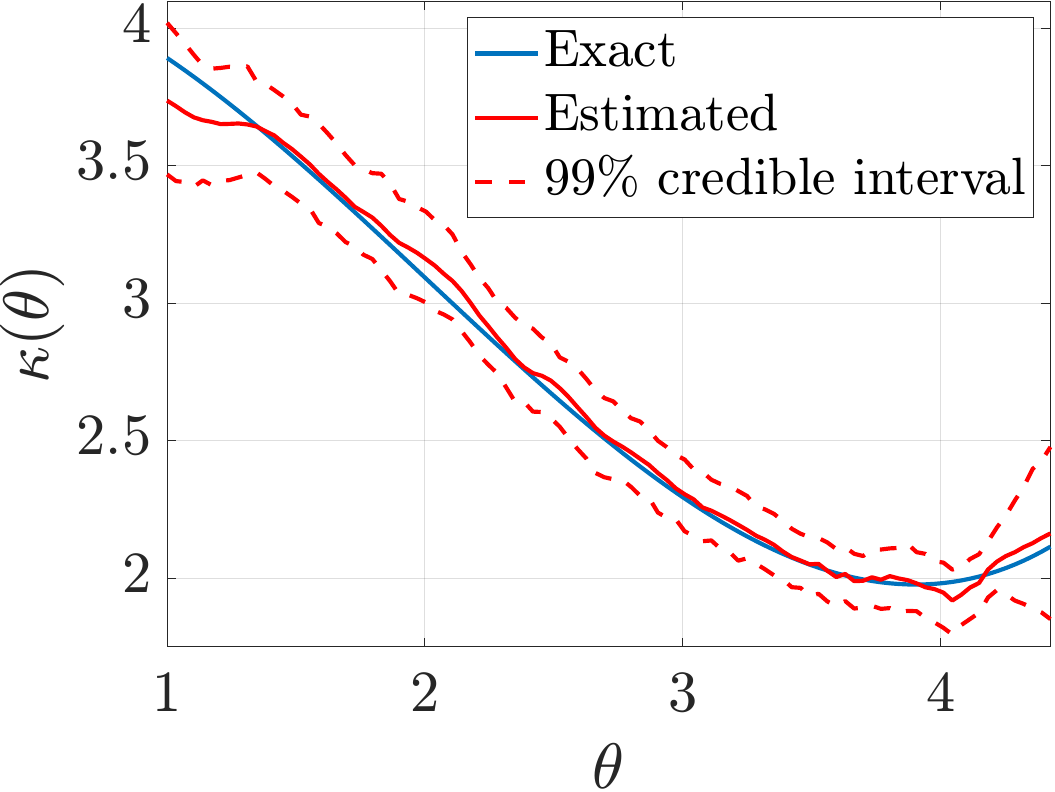}
                    \caption{$\gamma^2 = \num{2e-3}$.}
                    \label{figure-estimation-and-uq-gmrf-Qexact-gamma2e-3}
                \end{subfigure}
                \hfil
                \begin{subfigure}{0.45\linewidth}
                    \includegraphics[width=\linewidth]{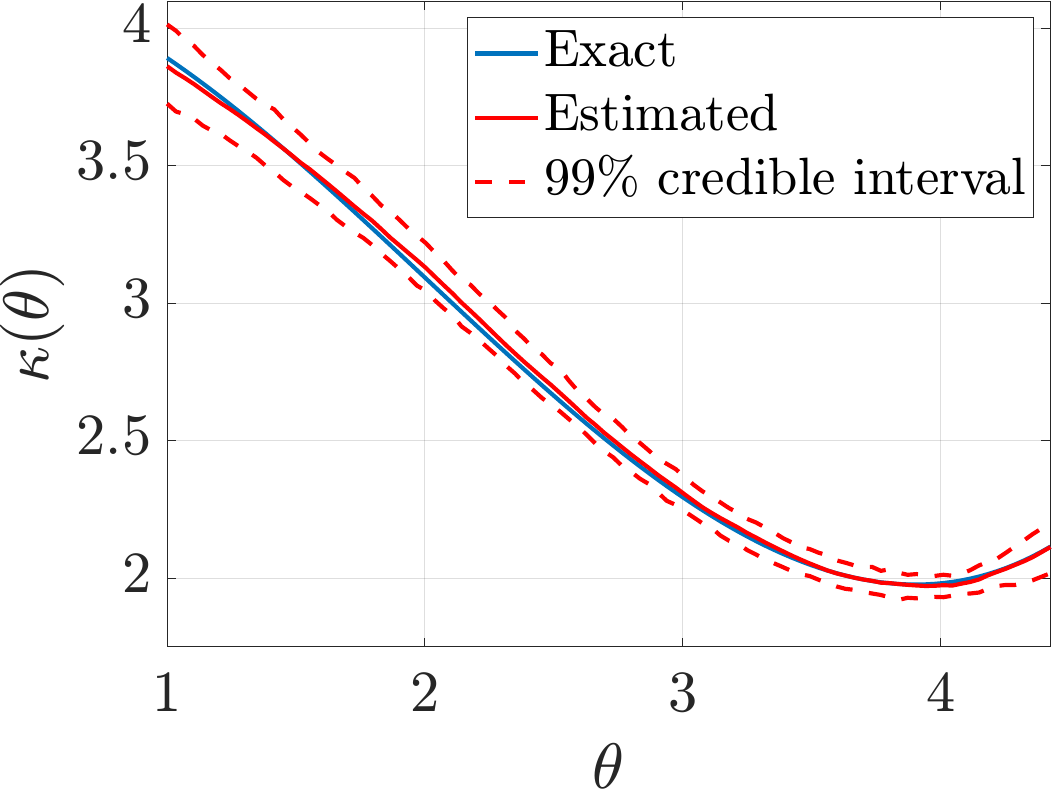}
                    \caption{$\gamma^2 = \num{2e-4}$.}
                    \label{figure-estimation-and-uq-gmrf-Qexact-gamma2e-4}
                \end{subfigure}
                \par\medskip
                \begin{subfigure}{0.45\linewidth}
                    \includegraphics[width=\linewidth]{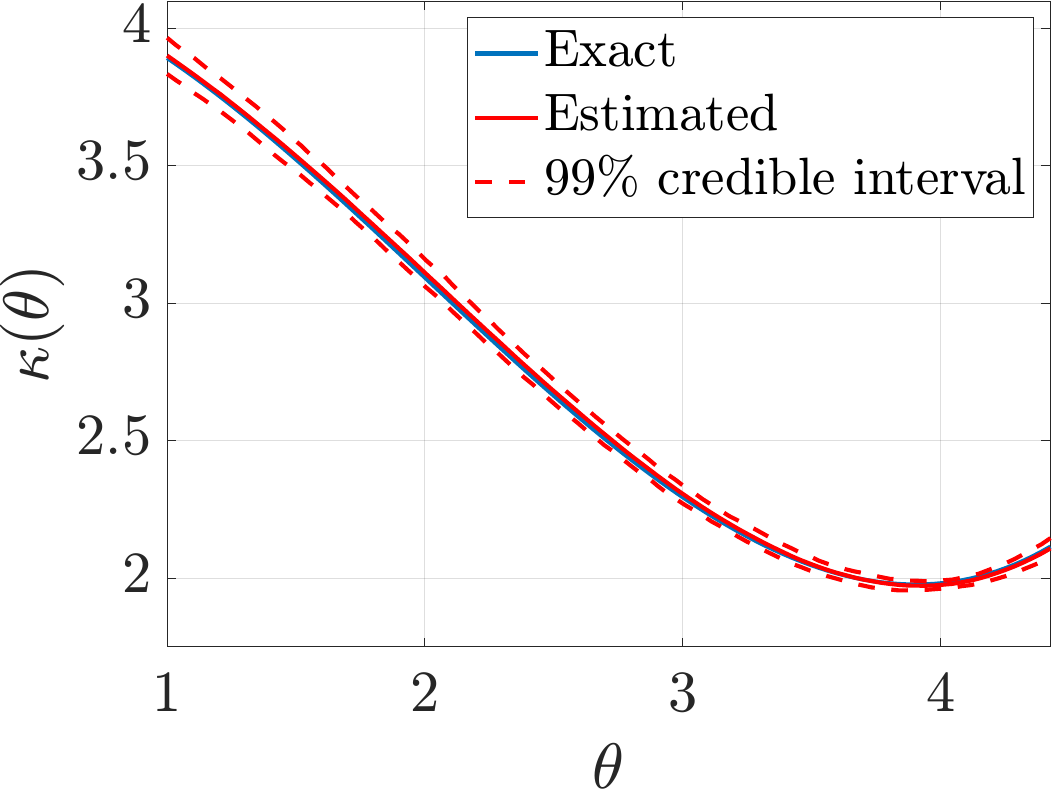}
                    \caption{$\gamma^2 = \num{2e-5}$.}
                    \label{figure-estimation-and-uq-gmrf-Qexact-gamma2e-5}
                \end{subfigure}
                \caption{Comparison between the exact and expected values of $\kappa(\theta)$ obtained with the piecewise linear functions model and the GMRF prior with $\bar{\bm{Q}} = \bm{Q_{\text{exact}}}$ and different values of $\gamma^2$.}
                \label{figure-estimation-and-uq-gmrf-Qexact}
            \end{figure}

            \autoref{figure-estimation-and-uq-gmrf-Qexact} shows the results obtained with $\bar{\bm{Q}} = \bm{Q_{\text{exact}}}$ and different values of $\gamma^2$.
            A total of \num{5e4} samples of state are considered.
            This number is larger compared to the one used in our third-degree polynomial model because we noticed that, when we use the piecewise-linear functions, the Markov chains require more samples to reach equilibrium.
            When we set $\gamma^2 = \num{2e-3}$ (\autoref{figure-estimation-and-uq-gmrf-Qexact-gamma2e-3}), this variance is one order of magnitude larger than the exact one.
            As a result, we notice more oscillations for the expected values of $\kappa$ and larger credible intervals.
            \autoref{figure-estimation-and-uq-gmrf-Qexact-gamma2e-4} shows the results obtained when we set $\gamma^2$ to the exact value of \num{2e-4}.
            Here we notice a very good agreement between the expected and exact values of $\kappa$, which also occurs because we set $\bar{\bm{Q}} = \bm{Q_{\text{exact}}}$.
            Additionally, we observe a smaller credible interval when compared to the previous case.
            Finally, \autoref{figure-estimation-and-uq-gmrf-Qexact-gamma2e-5} shows the results for $\gamma^2 = \num{2e-5}$. Since this value is one order of magnitude smaller than the exact one, we immediately notice that this configuration provides the smallest credible interval.
            Similar to the results shown by Figures \ref{figure-estimation-and-uq-values-normal1} and \ref{figure-estimation-and-uq-coeff-normal1}, care must be taken while assuming small variances.
            Although they provide small credible intervals, if the mean values used in the prior are not properly specified, the results might lead to wrong interpretations.

\begin{figure}
                \centering
                \begin{subfigure}{0.45\linewidth}
                    \includegraphics[width=\linewidth]{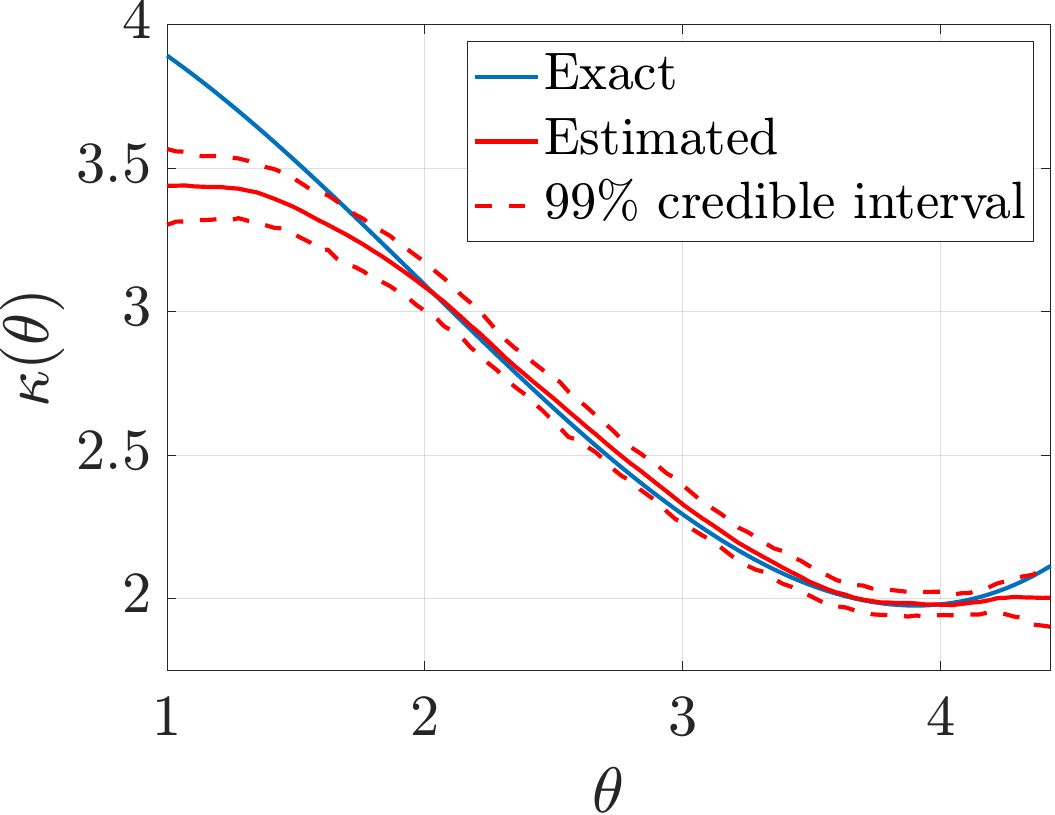}
                    \caption{$\bar{\bm{Q}} = \bm{0}$.}
                    \label{estimation-and-uq-gmrf-gammaExact-Q0}
                \end{subfigure}
                \hfil
                \begin{subfigure}{0.45\linewidth}
                    \includegraphics[width=\linewidth]{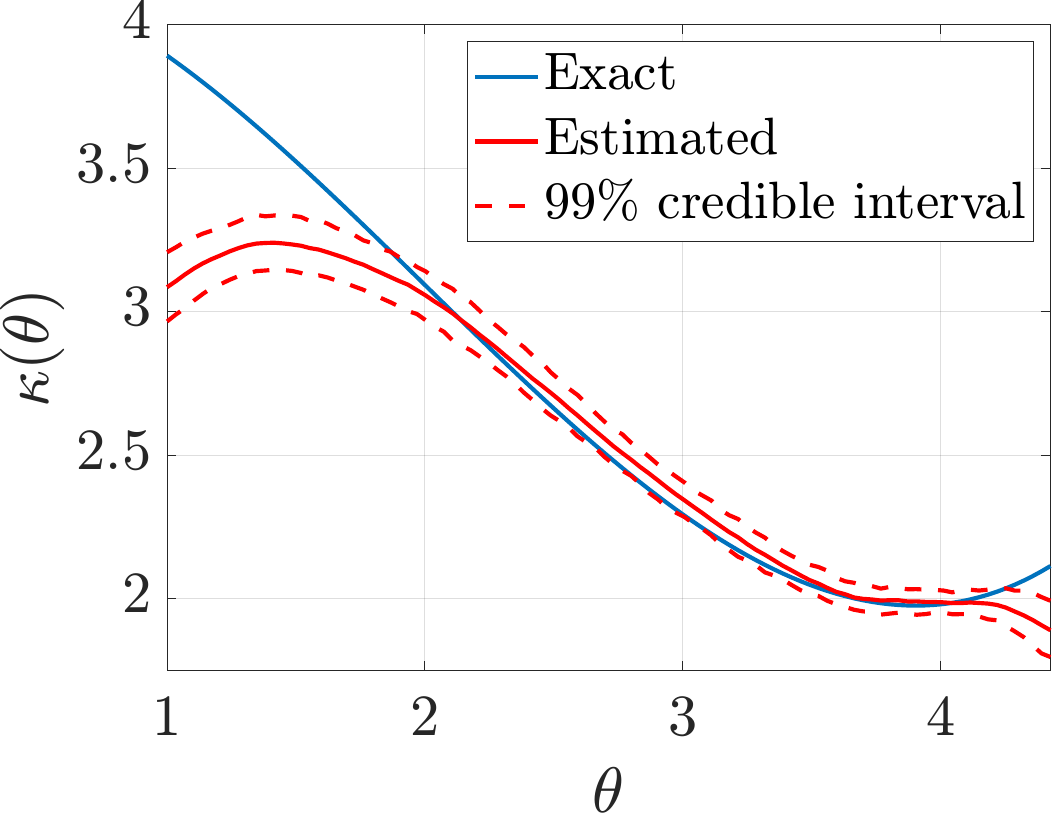}
                    \caption{$\bar{\bm{Q}} = - \bm{Q_{\text{exact}}}$.}
                    \label{estimation-and-uq-gmrf-gammaExact-Q-1}
                \end{subfigure}
                \caption{Comparison between the exact and expected values of $\kappa(\theta)$ obtained with the piecewise linear functions model and the GMRF prior with $\gamma^2 = \num{2e-4}$ and different values of $\bar{\bm{Q}}.$}
                \label{estimation-and-uq-gmrf-gammaExact}
            \end{figure}

            The results obtained for a fixed $\gamma^2 = \num{2e-4}$ and different values of $\bar{\bm{Q}}$ are illustrated by \autoref{estimation-and-uq-gmrf-gammaExact}.
            By setting $\bar{\bm{Q}} = \bm{0}$ (\autoref{estimation-and-uq-gmrf-gammaExact-Q0}), we represent a situation where we expect that the difference between consecutive neighbours in $\bm{P}$ is zero.
            The effects of the prior on the expected values are evident for values of $\theta$ close to 1 and 4.5.
            We observe for these regions that $\kappa$ tends to remain constant.
            Finally, \autoref{estimation-and-uq-gmrf-gammaExact-Q-1} illustrates the results obtained when we set $\bar{\bm{Q}} = - \bm{Q}_{\text{exact}}$.
            This figure shows that, when $\theta$ is close to 1, the variation of the exact values of $\kappa$ is negative, at the same time that the variation of the expected values is positive.
            Additionally, we see that when $\theta$ is close to 4.5, the exact values of $\kappa$ increase while the expected ones decrease.
            This behavior was already expected because we set $\bar{\bm{Q}} = - \bm{Q}_{\text{exact}}$.
            Nevertheless, for both results shown in Figure~\ref{estimation-and-uq-gmrf-gammaExact}, we see a good agreement between the exact and expected values of $\kappa$ for values of $\theta$ approximately between 2 and 4.
            This shows that, even if $\bar{\bm{Q}}$ does not precisely model the difference between consecutive elements in $\bar{\bm{P}}$, the prior still provides sufficiently precise expectations for the function $\kappa(\theta)$.

    \section{Conclusions}
    \label{section-conclusions}

    This work shows that the proposed Bayesian framework is able to provide the estimation and uncertainty quantification of a temperature-dependent thermal conductivity.
    The non-linearity of the physical problem raises questions regarding how to model the temperature dependence of the conductivity.
    Our study is conducted by selecting two models to represent the thermal conductivity as a function of the temperature: a third-degree polynomial and piecewise linear functions.
    The obtained results provide us insight about their respective advantages and disadvantages.
    
    Regarding the third-degree polynomial, we notice a greater practical value in selecting the parametrization in which the coefficients are estimated by first estimating values of the conductivity.
    This is due to the fact that, in contrast to the coefficients, the conductivities have a clear physical meaning.
    Hence, it is easier to define prior knowledge about the quantities of interest.
    This allows us to properly select reference values for the sensitivity analysis, thus obtaining insight about which quantities can be simultaneously estimated.
    Next to that, if we use a third-degree polynomial as our model, the estimation and uncertainty quantification of the conductivity is conducted via the inference of only 4 parameters.
    This allows us to select a relatively small number of samples of state for the Markov chains in order to reach equilibrium.
    The results obtained shows that, even when a non-informative improper uniform prior is considered, we can obtain a good agreement between the expected and exact values of the conductivity.
    Additionally, Gaussian priors with small standard deviations can be used to reduce the width of the credible intervals.
    Care must be taken while specifying the means and standard deviations, otherwise we can obtain a substantial disagreement between the exact and expected values of the conductivity.
    
    While using piecewise linear functions, the number of parameters to be inferred is equal to the number of points used to discretize the function $\kappa(\theta)$.
    In our example, it is assumed $N = 100$.
    This number is much larger than the number of parameters considered for the third-degree polynomial model.
    As a result, we immediately notice that the Markov chains require more samples in order to reach equilibrium.
    We notice that the GMRF is able to properly model the relation between consecutive values of $\kappa_n$.
    Hence, if there is prior knowledge about the relation between these values, the GMRF can be used to obtain an appropriate estimation and uncertainty quantification of $\kappa(\theta)$ when no assumption is made regarding the shape of this function.
    Likewise, this prior also allows us to obtain results that are not dependent on the initial guess.
    
    In conclusion, the choice of the model class to represent the thermal conductivity as a function of the temperature plays a crucial role.
    Our exploration of a third-degree polynomial and piecewise linear functions revealed distinct advantages and challenges associated with each approach.
    We strongly recommend aligning the chosen model with available prior information, emphasizing the need for careful consideration of prior choices and their impact on the estimation and uncertainty quantification of the constitutive model.
    Furthermore, exploring multiple models and comparing their performances can provide valuable insights into the underlying physical processes and enhance the robustness of the Bayesian framework.

    \section*{Acknowledgement}               
    This research was conducted as part of the DAMOCLES project within the EMDAIR program of the Eindhoven Artificial Intelligence Systems Institute (EAISI).
    
     \bibliographystyle{elsarticle-num}
     \bibliography{paper}
    
\end{document}